\documentclass{aa}     

\usepackage{graphicx}
\usepackage{txfonts}
\usepackage{natbib}
\usepackage{url}
\bibpunct{(}{)}{;}{a}{}{,}

\newcommand{\be}{\begin{equation}}
\newcommand{\ee}{\end{equation}}

\def\apjl{ApJL}
\def\apjs{ApJS}

\newcommand{\Mpc}{$h^{-1}$\thinspace Mpc}

\newcommand{\vmh}{h^{-1}\mathrm{Mpc} }

\begin{document}  

\title{Multimodality of rich clusters from the SDSS DR8
within the supercluster-void network.
} 

\author {M.~Einasto\inst{1} \and L.J.~Liivam\"agi\inst{1,2} 
\and E.~Tempel\inst{1,3} \and E.~Saar\inst{1,4} \and  J.~Vennik\inst{1}   
\and P.~Nurmi\inst{5}  \and M.~Gramann\inst{1}
\and J.~Einasto\inst{1,4,6}  \and E.~Tago\inst{1}
 \and  P.~Hein\"am\"aki\inst{5} \and A.~Ahvensalmi\inst{5}
\and V.J.~Mart\'{\i}nez\inst{7}
}

\institute{Tartu Observatory, 61602 T\~oravere, Estonia
\and
Institute of Physics, Tartu University, T\"ahe 4, 51010 Tartu, Estonia
\and 
National Institute of Chemical Physics and Biophysics, Tallinn 10143, Estonia
\and
Estonian Academy of Sciences,  EE-10130 Tallinn, Estonia
\and 
Tuorla Observatory, University of Turku, V\"ais\"al\"antie 20, Piikki\"o, Finland
\and 
ICRANet, Piazza della Repubblica 10, 65122 Pescara, Italy
\and 
Observatori Astron\`omic, Universitat de Val\`encia, Apartat
de Correus 22085, E-46071 Val\`encia, Spain
}

\authorrunning{M. Einasto et al. }

\offprints{M. Einasto}

\date{ Received   / Accepted   }

\titlerunning{Environment of clusters}

\abstract
{
The study of the properties of galaxy clusters and their environment 
gives us information about the formation and evolution
of galaxies, groups and clusters, and larger structures -- superclusters
of galaxies and the whole cosmic web.
}
{
We study the relations between the multimodality  
of galaxy clusters drawn from the SDSS DR8 
and the environment where they reside. As cluster environment we consider
the global luminosity density field, supercluster membership,
and supercluster morphology. 
}
{
We use 3D normal mixture modelling,
the Dressler-Shectman  test, and  the peculiar velocity of cluster main galaxies
as signatures of multimodality of clusters. We calculate the luminosity 
density field to study the environmental densities around clusters, and to find
superclusters where clusters reside. 
We determine the morphology of superclusters with the Minkowski functionals
and compare the properties of clusters in superclusters of different morphology. 
We apply principal component analysis to study the relations between the
multimodality parametres of clusters and their environment simultaneously. 
}
{Multimodal clusters reside in higher density environment than
unimodal clusters. Clusters in superclusters have higher probability 
to have substructure than isolated clusters.
The superclusters can be divided into two main morphological types, spiders
and filaments.  Clusters in superclusters of spider morphology 
have higher probabilities to have substructure
and larger peculiar velocities of their main galaxies 
than clusters in superclusters of filament morphology.
The most luminous clusters are located in the high-density
cores of rich superclusters. Five of seven most luminous clusters,
and five of seven most multimodal clusters 
reside in spider-type superclusters; four of seven most unimodal clusters reside
in filament-type superclusters. 
}
{ Our study shows the 
importance of the role of superclusters as high density environment which 
affects  the properties of galaxy systems in them.
}


\maketitle

\section{Introduction} 
\label{sect:intro} 

Most galaxies in the Universe are located in groups and clusters of galaxies, 
which themselves reside in larger systems -- in superclusters of galaxies or in 
filaments crossing underdense regions between superclusters 
\citep{1978MNRAS.185..357J,  zes82, 1983ARA&A..21..373O, 1986ApJ...302L...1D}. 
Cluster studies, in combination with the study of their environment are needed 
to understand the physics of clusters themselves, and  the evolution of 
structure in the Universe.

In the $\Lambda$CDM concordance cosmological model groups and clusters of 
galaxies and their filaments are created by density perturbations of scale up to 
32~\Mpc, and superclusters of galaxies by larger perturbations, up to 100~\Mpc\ 
\citep{2011A&A...534A.128E, 2011A&A...531A.149S}. Still larger perturbations 
modulate the richness of galaxy systems. Superclusters of galaxies are the 
largest density enhancements in the cosmic web. Studies of their properties and 
galaxy and cluster content are needed to understand the formation, evolution, 
and  properties of the large-scale structure and to compare cosmological models 
with observations \citep[][and references therein]{kbp02, 2007A&A...462..397E, 
e07, 2007JCAP...10..016H, 2009MNRAS.399...97A, 2011ApJ...736...51E, 
2011MNRAS.417.2938S, 2012arXiv1201.1382L}. 

The structures forming the cosmic web grow by hierarchical clustering driven by 
gravity \citep[see, e.g.,][and references therein]{2002PhRvD..65d7301L, 
loeb2008}. Galaxy clusters form at intersections of filaments, through them 
galaxies and galaxy groups merge with clusters. An indicator of former or 
ongoing mergers in groups and clusters of galaxies is their multimodality: the 
presence of a substructure (several galaxy associations within clusters), a 
large peculiar velocities of their main galaxies, and non-Gaussian velocity 
distribution of their galaxies \citep[][hereafter E12]{1993AJ....105.1596B, 
1996ApJS..104....1P, 1999A&A...343..733S, 2000A&A...354..761K, bur04, 
2006A&A...450....9F, 2008A&A...487...33B, 2012ApJ...746..139A, 
2012arXiv1201.3676H, 2012arXiv1202.4927E}. More references can be found in E12. 

Several studies have shown that richer and more luminous groups and clusters of 
galaxies from observations and simulations are located in a higher density 
environment \citep[][and references therein]{e2003a, e2003b, 
2005A&A...436...17E, 2011arXiv1109.4169C, 2011arXiv1111.1757P}. 
\citet{2002MNRAS.329L..47P} and \citet{2004ogci.conf...19P} showed that 
dynamically younger clusters are more strongly clustered than overall cluster 
population.

In this study we analyse the relations between the multimodality of rich 
clusters from the SDSS DR8  and the environment where they reside. We calculate 
the luminosity density field to trace the supercluster-void network, to define 
the values of the environmental density around clusters, and to determine 
superclusters of galaxies. For each cluster we find whether the cluster is 
located in a supercluster and study the relations between the properties of 
superclusters and clusters. We compare the properties of isolated clusters, and 
clusters in superclusters, and compare the properties of clusters in 
superclusters of different morphology, to understand whether the morphology of 
superclusters is also an important environmental factor in shaping the 
properties of  groups and clusters in superclusters. 

E12 analysed the substructure and velocity distributions of galaxies in the 
richest clusters from the SDSS DR8 with at least 50 member galaxies using a 
number of tests of different dimensionality. They showed that two most sensitive 
tests for the presence of substructure were 3D normal mixture modelling and the 
Dressler-Shectman (DS or $\Delta$) test \citep[see also comments in ][about the 
sensitivity of various tests]{1996ApJS..104....1P, 2012arXiv1201.3676H}. In this 
study we use the results of these two tests as an indicators of cluster 
substructure, and the peculiar velocities of the main galaxies in clusters. With 
principal component analysis we study the relation between the multimodality of 
clusters and their environment characterised by the values of the environmental 
density and supercluster luminosities. In Sect.~\ref{sect:data} we describe the 
data we used, in Sect.~\ref{sect:results} we give the results. We discuss the 
results and draw conclusions in Sect.~\ref{sect:discussion}.

We assume  the standard cosmological parametres: the Hubble parametre $H_0=100~ 
h$ km~s$^{-1}$ Mpc$^{-1}$, the matter density $\Omega_{\rm m} = 0.27$, and the 
dark energy density $\Omega_{\Lambda} = 0.73$.

\section{Data} 
\label{sect:data}

We use the MAIN galaxy sample of the 8th data release of the Sloan Digital Sky 
Survey \citep{2011ApJS..193...29A} with the apparent $r$ magnitudes $r \leq 
17.77$, and the redshifts $0.009 \leq z \leq 0.200$, in total 576493 galaxies. 
We corrected the redshifts of galaxies for the motion relative to the CMB and 
computed the co-moving distances \citep{mar03} of galaxies. The absolute 
magnitudes of galaxies were determined in the $r$-band ($M_r$) with the $k$-
corrections for the SDSS galaxies, calculated using the KCORRECT algorithm 
\citep{2007AJ....133..734B}. In addition, we applied evolution corrections, 
using the luminosity evolution model of \citet{blanton03b}. The magnitudes 
correspond to the rest-frame at the redshift $z=0$. The details about data 
reduction and the description of the group catalogue can be found in 
\citet{2011arXiv1112.4648T}.

We determine groups of galaxies using the Friends-of-Friends cluster analysis 
method introduced in cosmology by \citet{tg76,zes82,hg82}. A galaxy belongs to a 
group of galaxies if this galaxy has at least one group member galaxy closer 
than a linking length. In a flux-limited sample the density of galaxies slowly 
decreases with distance. To take this selection effect into account properly 
when constructing a group catalogue from a flux-limited sample, we rescaled the 
linking length  with distance, calibrating the scaling relation by observed 
groups \citep[see ][for details]{tago08, 2010A&A...514A.102T}. As a result, the 
maximum sizes in the sky projection and the velocity dispersions of our groups 
are similar at all distances. 

We use in this study systems from the group catalogue with at least 50 member 
galaxies analysed for substructure in E12. These clusters are chosen from the 
distance interval 120~\Mpc\ $\le D \le $ 340~\Mpc\ (the redshift range $0.04 < z 
< 0.12$) where the selection effects are the smallest \citep[we discuss the 
selection effects in detail in E12 and in ][]{2010A&A...514A.102T}. This sample 
of 109  clusters includes all clusters from the SDSS DR8 with at least 50 member 
galaxies in this distance interval. E12 showed that more than 80\% of clusters 
in this sample demonstrate a signs of multimodality according to several 3D, 2D, 
and 1D tests: the presence of multiple components, large probabilities to have a 
substructure, and the deviations of galaxy velocity distributions in clusters 
from Gaussianity. The larger the dimensionality of the test, the more sensitive 
it is to the presence of substructure in clusters (for details we refer to E12). 
In this study we use the results of two 3D test to characterise the 
multimodality in clusters: the 3D normal mixture modelling and the Dressler-
Shectman (DS) test. We describe these tests in Appendix~\ref{sect:multi}. In 
addition, we use the peculiar velocity of the main galaxies, $V_{\mathrm{pec}}$. 
In the group catalogue the main galaxy of a group is defined as the most 
luminous galaxy in the $r$-band. We use this definition also in the present 
paper.

We calculate the galaxy luminosity density field to reconstruct the underlying 
luminosity distribution, and to find the  environmental density around clusters. 
Environmental densities are important for understanding the influence of the 
local and/or global environment on cluster properties. Three smoothing lengths 
are used for environmental densities around clusters, 4, 8, and 16~\Mpc\ to 
characterise environment at scales around clusters from cluster local 
surroundings to supercluster scales. For details we refer to 
Appendix~\ref{sect:DF} and to \citet{2011arXiv1112.4648T}.

To determine supercluster membership for clusters, we first found superclusters 
(extended systems of galaxies) in the luminosity density field at smoothing 
length 8~\Mpc. We created a set of density contours by choosing a density 
threshold and defined connected volumes above a certain density threshold as 
superclusters. In order to choose proper density levels to determine individual 
superclusters, we analysed the density field superclusters at a series of 
density levels. As a result we used the density level $D = 5.0$ (in units of 
mean density, $\ell_{\mathrm{mean}}$ = 1.65$\cdot10^{-2}$ $\frac{10^{10} h^{-2} 
L_\odot}{(\vmh)^3})$ to determine individual superclusters. At this density 
level superclusters in the richest chains of superclusters in the volume under 
study  still form separate systems; at lower density levels they join into huge 
percolating systems. At higher threshold density levels superclusters are 
smaller and their number decreases. 

Superclusters are characterised by their total luminosity, richness, and 
morphology, determined with Minkowski functionals. The total luminosity of the 
superclusters $L_{\mathrm{scl}}$ is calculated as  the sum of weighted galaxy 
luminosities: \begin{equation} L_{\mathrm{scl}} = \sum_{\mathrm{gal} \in 
\mathrm{scl}} W_L (d_{\mathrm{gal}}) L_{\mathrm{gal}}. \label{eq:wlum} 
\end{equation} Here the $W_L(d_{\mathrm{gal}})$ is the distance-dependent weight 
of a galaxy (the ratio of the expected total luminosity to the luminosity within 
the visibility window). The description of the supercluster catalogues is given 
in \citet{2010arXiv1012.1989J} and in Liivam\"agi et al. (in preparation, DR8 
catalogue). 


\begin{figure*}[ht]
\centering
\resizebox{0.95\textwidth}{!}{\includegraphics[angle=0]{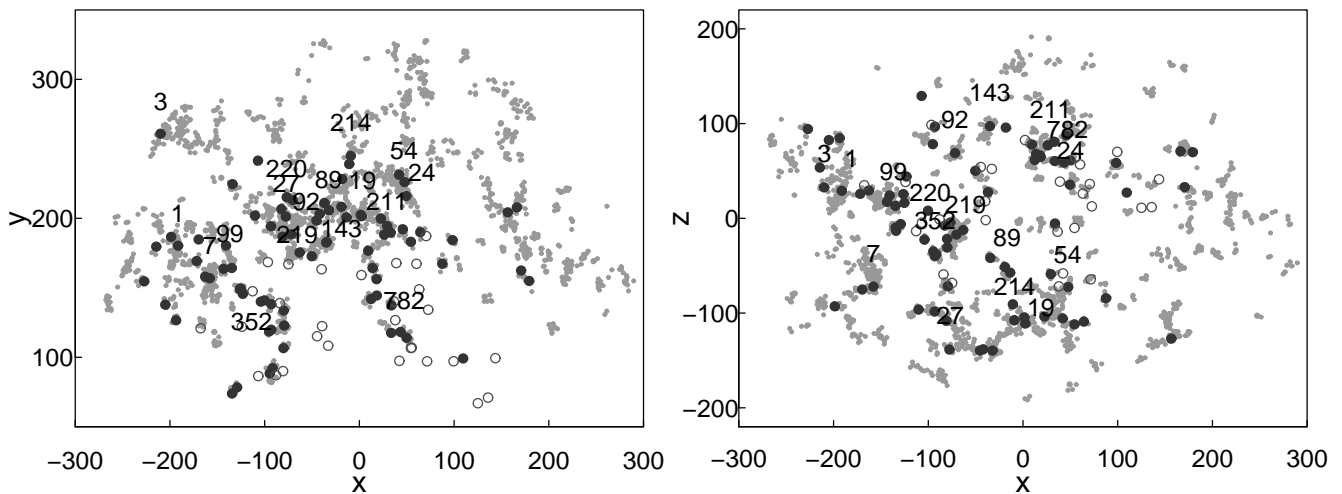}}
\caption{
Distribution  of groups with at least
four member galaxies in superclusters in x, y, and z coordinates (in $h^{-1}$ Mpc,
grey dots).  
Black filled circles denote clusters with at least 50 member galaxies in 
superclusters, dark grey empty circles those  clusters with at least 50 member galaxies 
which are not located in superclusters.
Numbers are ID numbers of superclusters with at least 500 member galaxies.
}
\label{fig:richxyz}
\end{figure*}

The supercluster morphology is fully characterised by the four Minkowski 
functionals \mbox{$V_0$--$V_3$}. For a given surface the four Minkowski 
functionals (from the first to the fourth) are proportional to the enclosed 
volume $V$, the area of the surface $S$, the integrated mean curvature $C$, and 
the integrated Gaussian curvature $\chi$ \citep{sah98, mar03, sss04, saar06, 
saar09}. We give formulaes in Appendix~\ref{sect:MF}.

The overall morphology of a supercluster is described by the  shapefinders $K_1$ 
(planarity) and $K_2$ (filamentarity), and their ratio, $K_1$/$K_2$ (the shape 
parametre), calculated using the first three Minkowski functionals. They contain 
information both about the sizes of superclusters and about their outer shape. 
The smaller the shape parametre, the more elongated a supercluster is.

The maximum value of the fourth Minkowski functional $V_3$ (the clumpiness) 
characterises the inner structure of the superclusters and gives the number of 
isolated clumps, the number of void bubbles, and the number of tunnels (voids 
open from both sides) in the region \citep[see, e.g.][]{saar06}. The larger the 
value of $V_3$, the more complicated the inner morphology of a supercluster is; 
superclusters may be clumpy, and they also may have holes or tunnels in them 
\citep{e07, 2011ApJ...736...51E}. Superclusters show large morphological variety 
in which \citet{2011A&A...532A...5E} determined four main morphological types: 
spiders, multispiders, filaments, and multibranching filaments. Spiders and 
multispiders are systems of one or several high-density clumps with a number of 
outgoing filaments connecting them. The Local Supercluster is an example of a 
typical poor spider. Filaments and multibranching filaments are superclusters 
with filament-like main body which connects clusters. An example of an 
exceptionally rich and dense multibranching filament is the richest supercluster 
in the Sloan Great Wall \citep{e07, 2011ApJ...736...51E}. For simplicity, in 
this study we classify superclusters as spiders and filaments. 

Data about clusters and superclusters are given in online 
Tables~\ref{tab:cldata1} and  ~\ref{tab:scl}. We cross-identify groups with 
Abell clusters (Table~\ref{tab:cldata1}). We consider a group identified with an 
Abell cluster, if the distance between their centres is smaller than at least 
the linear radius of one of the clusters, and the distance between their centres 
in the radial (line-of-sight) direction is less than 600  $km~s^{-1}$  (an 
empirical value). In some cases one group can be identified with more than one 
Abell cluster and vice versa (for details we refer to E12). In 
Table~\ref{tab:scl} we give to superclusters the ID number from \citet{e2001} 
catalogue if there is at least one Abell cluster in common between this 
catalogue and the present supercluster sample. A common cluster does not always 
mean that superclusters can be fully identified with each other. A number of 
superclusters from E01 are split between several superclusters in our present 
catalogue, an examples of such systems are SCl~019 and SCl~054, which both 
belong to SCl~111 in \citet{e2001} catalogue.

\section{Results}
\label{sect:results} 

\subsection{The large-scale environment of clusters}
\label{sect:lss} 

\begin{figure}[ht]
\centering
\resizebox{0.45\textwidth}{!}{\includegraphics[angle=0]{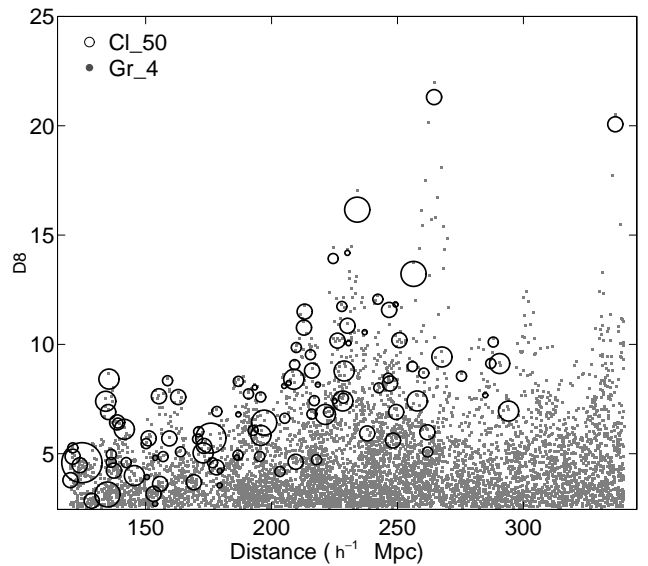}}
\caption{
Global densities at smoothing length 8 \Mpc\ (in units of mean density) around groups
and clusters vs. their distance. Black circles denote clusters
with at least 50 member galaxies, the size of circles is proportional to the
number of components found by 3D normal mixture modelling. Grey dots denote groups 
with 4--49 member galaxies.  
}
\label{fig:dendist}
\end{figure}

To study the distribution of clusters in the
supercluster-void network we present in Figure~\ref{fig:richxyz} 
the distribution  of clusters with at least four 
member galaxies in superclusters, and the distribution of isolated clusters with 
at least 50 member galaxies in cartesian coordinates $x$, $y$, and $z$ defined as in
\citet{2007ApJ...658..898P} and in \citet{2010arXiv1012.1989J}:
\begin{equation}
\begin{array}{l}
    x = -d \sin\lambda, \nonumber\\[3pt]
    y = d \cos\lambda \cos \eta,\\[3pt]
    z = d \cos\lambda \sin \eta,\nonumber
\end{array}
\label{eq:xyz}
\end{equation}
where $d$ is the comoving distance, and $\lambda$ and $\eta$ are the SDSS 
survey coordinates. In Fig.~\ref{fig:dendist}
we plot the values of the environmental density around groups with 
at least 4 member galaxies at smoothing 
length 8 \Mpc\ vs. the distance of groups. In this figure circles represent
clusters with at least 50 member galaxies. 
The size of circles is proportional to the number
of components in clusters determined with the 3D normal mixture modelling. 

Figure~\ref{fig:richxyz} (left panel) shows that at the smallest distances from 
us (at low $y$ values, up to distances approximately 180~\Mpc) the sample 
crosses the void region. This is the void between the nearby rich superclusters 
\citep[the Hercules supercluster and other systems, the detailed description of 
the large-scale distribution of superclusters in this region was given 
in][]{2011A&A...532A...5E}. Groups and clusters form two filaments of poor 
superclusters and isolated clusters crossing this void. The richest 
superclusters in these filaments are SCl~352 and SCl~782 (we identify 
supercluster members among clusters in Sect.~\ref{sect:scl}). The density 
distribution in Fig.~\ref{fig:dendist} shows that even the maximal values of the 
environmental densities in this region are low, up to $D8 \approx 5$, in the 
density peaks in filaments at the locations of superclusters $D8 < 8$ (in units 
of the mean density). Figures~\ref{fig:dendist} and ~\ref{fig:richxyz} shows 
that rich clusters in superclusters mark the peaks in the density distribution, 
isolated clusters are located at lower densities in filaments. The sizes of 
symbols in Fig.~\ref{fig:dendist} show that among these clusters there are both 
multicomponent and one-component clusters. 

At distances between 180 \Mpc\ and 270 \Mpc\ the SDSS survey crosses systems of 
rich superclusters. The richest superclusters in these systems are SCl~027  and 
SCl~019 in the Sloan Great Wall, SCl~211 (the Ursa  Major supercluster) in 
another chain of superclusters, and SCl~099 (the Corona Borealis supercluster) 
and SCl~001 in the dominant supercluster plane at the intersection of the 
supercluster chains \citep[for details see][]{1997A&AS..123..119E, 
2011A&A...532A...5E}. The values of the environmental densities $D8 < 8$ in the 
foreground of rich superclusters at distances less than 200~\Mpc. At larger 
distances, in rich superclusters  the maximal values of the environmental 
densities are much higher than in the void region behind them. Again rich 
clusters mark the high density peaks in the density field 
(Fig.~\ref{fig:dendist}). Some rich clusters are located in the cores of rich 
superclusters with the highest values of the environmental densities,  $D8 > 10$ 
\citep{e07b, tempel09, 2011A&A...529A..53T}. The environmental density is the 
largest ($D8 = 21.3$) in the supercluster SCl~001, around rich cluster 29587 
(Abell cluster A2142). At still larger distances the SDSS sample reaches the 
void region behind the Sloan Great Wall and other rich superclusters, and the 
values of the environmental densities are lower again. The farthest rich cluster 
in our sample belong to the rich supercluster SCl~003 behind this void at a 
distance of 336~\Mpc. Here the value of the environmental density is also very 
high, $D8 = 20.1$.

Figure~\ref{fig:dendist} shows that the lowest values of the environmental 
densities around rich clusters slightly increase with distance. This is due to 
the use of groups with 50 and more members. Due to the flux-limited sample, 
the groups with the same richness are also brighter further away. In E12 we showed 
that the richness of rich clusters in our sample does not depend on distance, 
therefore our sample of clusters is not strongly affected by this selection 
effects. When comparing the environmental densities around clusters in some 
cases we shall use two distance intervals, to analyse densities and the 
properties of clusters in void region and in supercluster region separately, and 
to minimise the influence of this selection effect.
  
A visual inspection of  Fig.~\ref{fig:dendist} shows that both multicomponent 
and one-component clusters are located in all density peaks. 
Next we analyse the values of the environmental densities around clusters in more 
detail.

\begin{figure*}[ht]
\centering
\resizebox{0.95\textwidth}{!}{\includegraphics[angle=0]{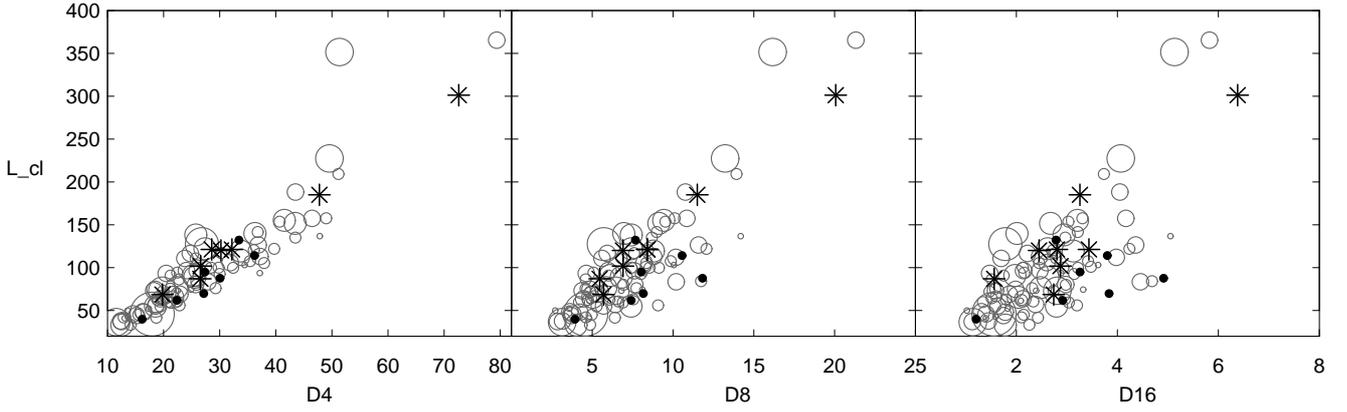}}
\caption{
From left to right: cluster luminosities (in $10^{10} h^{-2} L_{\sun}$)
vs. their environmental densities at the smoothing lengths 
4, 8, and 16  \Mpc\ (in units of mean density) (grey circles). 
The size of circles is proportional to the
number of components found by 3D normal mixture modelling. 
Stars denote most multimodal clusters,
and filled circles denote most unimodal clusters as described in the text.
}
\label{fig:lumden}
\end{figure*}

We plot cluster luminosities vs. environmental densities at three smoothing 
lengths in Fig.~\ref{fig:lumden}, and search  for the pairwise correlations 
between the  parametres of clusters and the environmental densities around them 
with the Pearson's correlation test (Table~\ref{tab:corr}). In 
Fig.~\ref{fig:lumden} we mark those clusters which are unimodal according to all 
the tests applied in E12 with filled circles, and those which are  multimodal 
with stars (we discuss them in detail in Sect.~\ref{sect:selcl}).

We exclude from this analysis the cluster Gr1573 near the edge of the survey
for which environmental densities cannot be determined reliably
($D = -999$ in Table~\ref{tab:cldata1}). 

At the smoothing length 4~\Mpc\ the luminosity density is determined by cluster 
members, and galaxies and galaxy systems in the close neighbourhood of clusters. 
Therefore the correlation between the luminosities of clusters and environmental 
densities is strong (the correlation coefficient is $r = 0.91$ with very high 
statistical significance, Table~\ref{tab:corr}). Figure~\ref{fig:lumden} shows 
that densities around clusters of the same luminosity may differ up to 1.5 - 2 
times depending on the systems around clusters.  At this smoothing length there 
is no statistically significant correlation between the number of components in 
clusters and the environmental density around clusters. The statistical 
significance to have  substructure in clusters according to the DS test 
$p_{\mathrm{\Delta}}$ is weakly anticorrelated with the environmental density. 
Small $p_{\mathrm{\Delta}}$ values show higher significance of having 
substructure, therefore this test shows that there is a tendency that clusters 
with higher probabilities of having a substructure reside in a higher density 
environments. However, Table~\ref{tab:corr} shows that the statistical 
significance of this result is low. Figure~\ref{fig:lumden} shows that one of 
the most luminous clusters have relatively low density local environment around 
it (Gr34727 in the supercluster SCl~7, Table~\ref{tab:lumcl}).

At the smoothing length 8~\Mpc\ (Fig.~\ref{fig:lumden}, middle panel) the 
scatter of the relation between cluster luminosities and environmental densities 
increases -- the difference between the environmental densities around clusters 
of the same luminosity increases and the correlations between the cluster 
luminosities and environmental density become weaker. The scatter is espacially 
large at densities $D > 8$, in the cores of rich superclusters, where both high-
and low-luminosity clusters reside. All most luminous clusters are located in 
supercluster cores. In poor superclusters environmental densities are lower. At 
the largest smoothing length, 16~\Mpc\ which characterises large scale 
supercluster environment around clusters the scatter of the relation between 
cluster luminosities and environmental densities increases and the correlations 
between the clusters luminosities and environmental density become weaker. The 
number of components in clusters at large smoothing lengths is not correlated 
with the environmental density around clusters, the correlation between the 
probability to have substructure in clusters and environmental density also 
becomes weaker. The correlations between the peculiar velocities of the main 
galaxies in clusters and environmental density  are statistically highly 
significant -- clusters in higher density environments have larger peculiar 
velocities of the main galaxies. The correlation coefficients are not large, 
from 0.24 at smoothing length 4 \Mpc\ to 0.18 at smoothing length 16 \Mpc.  The 
correlations between the  number of galaxies in clusters and the environmental 
density of clusters are statistically highly significant (Table~\ref{tab:corr}). 
The correlations are not very strong, with the correlation coefficient $r 
\approx 0.5$, being stronger at small smoothing length and weaker at large 
smoothing length. In addition, the larger the smoothing length, the stronger are 
the correlations between the luminosity of superclusters and environmental 
density -- richer and more luminous superclusters have also higher environmental 
densities, as found earlier by \citet{2007A&A...462..397E}.

\begin{table}[ht]
\caption{Correlations between the environmental density around clusters,
and cluster parametres.}
\begin{tabular}{lcccccc} 
\hline\hline 
$\mathrm{Parametre}$   & \multicolumn{2}{c}{D4} &    \multicolumn{2}{c} {D8} & \multicolumn{2}{c} {D16}   \\  
\hline
                              & r & p & r & p &  r & p   \\  
\hline   
$L_{\mathrm{cl}}$             &  0.91    & 1e-16      &  0.85   &     1e-16    & 0.70 &  1e-16\\
$N_{\mathrm{gal}}$            & 0.59     & 2.7e-11    &  0.54   &  1.8e-09   & 0.42  & 6.2e-06 \\
\\
$N_{\mathrm{comp}}$         &   0.06   & 0.55   &  0.01    &   0.91    & -0.04   &  0.65 \\
$V_{\mathrm{pec}}$            &   0.24   & 0.01   &  0.19    &  0.04     & 0.18    &  0.07 \\
$p_{\mathrm{\Delta}}$                   &  -0.14   & 0.16   &  -0.08  &  0.39      & -0.05   &  0.57 \\
\\                                                                               
$L_{\mathrm{scl}}$             &  0.49    & 7.2e-08      &  0.63   &     4.6e-13  & 0.73 &  2e-16\\
\hline            
                
\label{tab:corr}  
\end{tabular}\\
\tablefoot{ 
Pearson's correlation coefficients $r$ and their $p$-values for correlations
between the environmental densities $D4$, $D8$, and $D16$ around clusters,
and the number of 3D components, peculiar velocities of main 
galaxies, the significance to have substructure according to the DS test, $p_{\mathrm{\Delta}}$,  
the luminosity of clusters, $L_{\mathrm{cl}}$, the 
numbers of galaxies, $N_{\mathrm{gal}}$, and the total luminosity of superclusters.
}         
\end{table}
          
\begin{figure*}[ht]
\centering
\resizebox{0.9\textwidth}{!}{\includegraphics[angle=0]{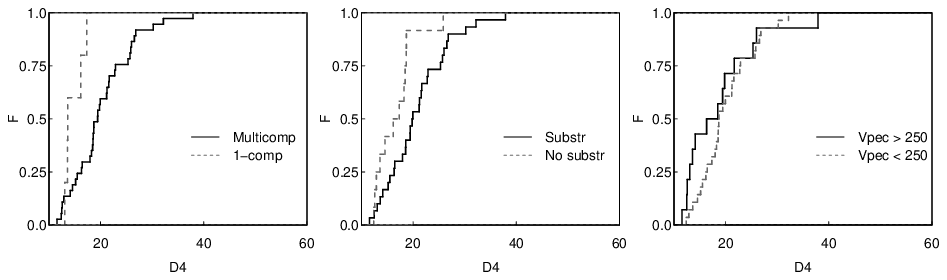}}
\resizebox{0.9\textwidth}{!}{\includegraphics[angle=0]{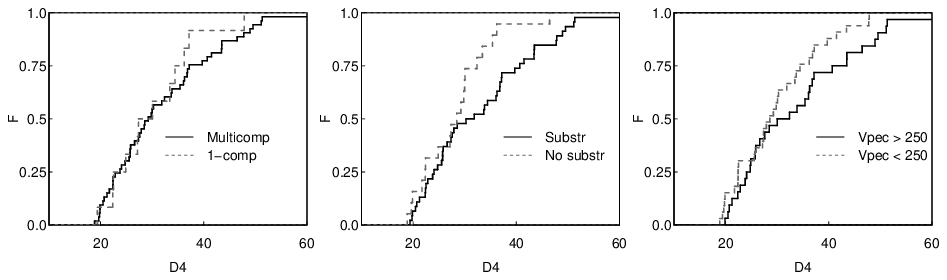}}
\caption{
Cumulative distributions of the values of the environmental densities around clusters
for the smoothing length 4~\Mpc. Solid black line denote densities around
 multicomponent clusters
(left panel), clusters with significant substructure (middle panel), and
clusters with the peculiar velocities of their main galaxies larger than 250 $km~s^{-1}$
(right panel). Dashed grey line denote densities around
 one-component clusters
(left panel), clusters without  significant substructure (middle panel), and
clusters with the peculiar velocities of their main galaxies smaller than 250 $km~s^{-1}$ 
(right panel). Upper row -- distance interval 120~\Mpc\ $\le D \le $ 180~\Mpc,
lower row -- distance interval 180~\Mpc\ $\le D \le $ 300~\Mpc.
}
\label{fig:d4d}
\end{figure*}

Figure~\ref{fig:d4d} shows the cumulative distributions of the values of the 
environmental densities at smoothing length 4 \Mpc, at which the correlations 
between the environmental density and multimodality parametres were the 
strongest, for two distance intervals: 120~\Mpc\ $\le D \le $ 180~\Mpc\ (upper 
row), and 180~\Mpc\ $\le D \le $ 300~\Mpc\ (lower row). There are 42 clusters in 
the closer distance interval, and 65 clusters in the farther distance interval. 
We show the cumulative distributions of densities around clusters divided into 
populations according to the different indicators of multimodality: 
multicomponent and one- component clusters according to the 3D normal mixture 
modelling, clusters with and without significant substructure according to the 
DS test ($p_{\mathrm{\Delta}} <= 0.05$, and $p_{\mathrm{\Delta}} > 0.05$, 
correspondingly), and clusters with small and large peculiar velocities of their 
main galaxies. E12 showed that, approximately, the peculiar velocity limit 
between these two populations is of about 250 $km~s^{-1}$. Figure~\ref{fig:d4d} 
shows that densities around multicomponent clusters and clusters with 
significant substructure  have higher values than densities around one-component 
clusters without significant substructure. The differences between the densities 
around clusters with small and large peculiar velocities of their main galaxies 
in the void region (120~\Mpc\ $\le D \le $ 180~\Mpc) are small, in the farther 
region in superclusters clusters with large peculiar velocities of their main 
galaxies have higher environmental densities around them than clusters with 
small values of the peculiar velocities of their main galaxies.

\begin{table}[ht]
\caption{Results of the principal component analysis for the multimodality 
and environmental parametres of clusters.
}
\begin{tabular}{lrrrrrr} 
\hline\hline 
                &       PC1 & PC2 & PC3 & PC4 & PC5 & PC6  \\
\hline
$N_{\mathrm{comp}}$     & 0.187 & -0.603 &  0.164 &  0.153 & -0.736 & -0.087 \\ 
$\log(V_{\mathrm{pec}})$  & 0.160 & -0.157 & -0.969 & -0.068 & -0.064 &  0.026 \\ 
$p_{\mathrm{\Delta}}$               & 0.272 & -0.504 &  0.148 & -0.703 &  0.380 & -0.094 \\ 
$\log(N_{\mathrm{gal}})$  & 0.512 & -0.247 &  0.070 &  0.555 &  0.412 &  0.437 \\ 
$\log(D8)$                & 0.612 &  0.312 &  0.020 &  0.122 &  0.016 & -0.714 \\ 
$\log(L_{\mathrm{scl}})$  & 0.475 &  0.445 &  0.073 & -0.390 & -0.371 &  0.528 \\ 
\hline
\multicolumn{3}{l}{Importance of components} && \\ 
\hline
                 &  PC1 & PC2 & PC3 & PC4 & PC5 & PC6  \\
St. dev.     & 1.397 & 1.183 & 0.983 & 0.870 & 0.833 & 0.481 \\
Prop. of var. & 0.325 & 0.233 & 0.161 & 0.126 & 0.116 & 0.039 \\
Cum. prop.  & 0.325 & 0.559 & 0.720 & 0.846 & 0.961 & 1.000 \\
\hline

\label{tab:pca}  
\end{tabular}\\
\tablefoot{ 
St. dev. denotes  standard deviation, Prop. of var. denotes proportion of 
variance, and Cum. prop. denotes cumulative proportion.
}
\end{table}

The relations between the parametres of clusters, the indicators of 
substructure, and the environmental parametres of clusters can be studied 
simultaneously with the principal component analysis (PCA). The PCA transforms 
the data to a new coordinate system, where the greatest variance by any 
projection of the data lies along the first coordinate (the first principal 
component), the second greatest variance -- along the second coordinate, and so 
on. There are as many principal components as there are parametres, but often 
only the first few are needed to explain most of the total variation. The 
principal components PC$i$  ($i \in \mathbb{N}$, $i \leq N_{\mathrm{tot}}$) are 
linear combinations of the original parametres:

\begin{equation}\label{eq:pc}
 PCi = \sum_{k=1}^{N_{\mathrm{tot}}} a(k)_{i} V_{k},
\end{equation}
where $-1 \leq a(k)_i \leq 1$ are the coefficients of the linear transformation, 
$V_k$ are the original parametres and $N_{\mathrm{tot}}$ is the number of the 
original parametres. In the analysis the parametres are  standardised -- they 
are centred on their means, $ V_{k} - \overline{V_{k}}$, and normalised, divided 
by their standard deviations, $\sigma( V_{k})$. E12 used PCA to analyse the 
relations between the multimodality parametres of clusters and their physical 
properties. We refer to \citet{2011A&A...535A..36E} for the references about 
applications of the PCA in astronomy.

\begin{figure}[ht]
\centering
\resizebox{0.45\textwidth}{!}{\includegraphics[angle=0]{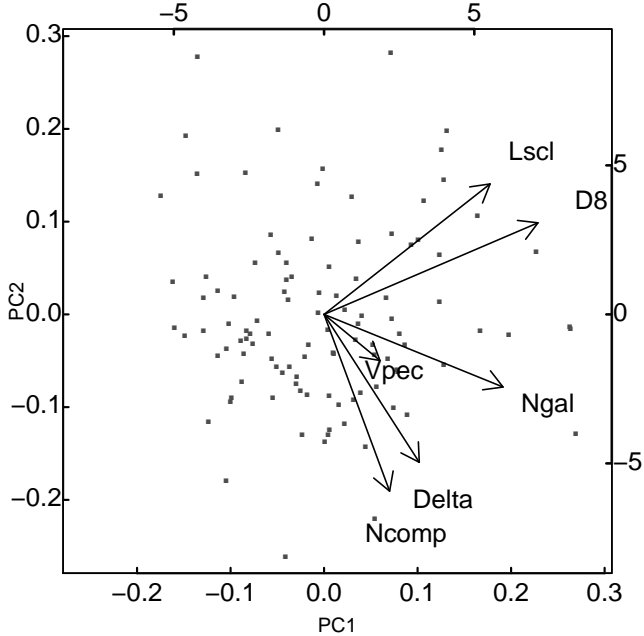}}
\caption{
Biplot of the principal component analysis  with the multimodality indicators  
and the environmental parametres of clusters, as described in the text. 
}
\label{fig:pcasubenv}
\end{figure}

We include into the calculations the number of components as determined with the 
3D normal mixture modelling, $p_{\mathrm{\Delta}}$ showing the probability to 
have substructure according to the DS test, the peculiar velocity of the main 
galaxy in clusters, $V_{\mathrm{pec}}$, the number of galaxies in clusters, 
$N_{\mathrm{gal}}$, the environmental density around clusters with smoothing 
length 8 \Mpc, $D8$ (environmental densities are correlated, therefore we 
include only one of them), and the luminosity of a supercluster where a cluster 
resides, $L_{\mathrm{scl}}$. We use $1 - p_{\mathrm{\Delta}}$ since larger 
values of $1 - p_{\mathrm{\Delta}}$ suggest a higher probability to have 
substructure, therefore the arrows in biplot corresponding to the number of the 
components point towards the same direction as the arrows corresponding to the 
DS test. We use logarithms of the peculiar velocities of main galaxies and 
environmental parametres. Figure~\ref{fig:pcasubenv} and Table~\ref{tab:pca} 
show the results of this analysis.

Table~\ref{tab:pca} shows that the coefficients of the first principal component 
are the largest for the environmental density around clusters, for the number of 
galaxies in clusters, and for the total luminosity of superclusters. This shows 
that richer clusters are located in a higher density environment, and richer 
superclusters have higher environmental densities in them \citep{e07b, 
2007A&A...462..397E}, as also shown with the analysis above. In the biplot 
showing the results of the PCA (Fig.~\ref{fig:pcasubenv}) the arrows 
corresponding to the tests about substructure and arrows corresponding to the 
other parametres of clusters are not pointed into the same direction. This 
suggests that the correlations between substructure parametres and the 
environment of clusters are not strong, as also the correlation calculations 
showed. In Fig.~\ref{fig:pcasubenv} the arrow corresponding to the peculiar 
velocity of the main galaxies in clusters is pointed approximately into the same 
direction as the arrow for  richness of clusters, showing that these velocities 
are larger in richer clusters. The length of the arrows and coefficients in 
Table~\ref{tab:pca} show that the importance of the peculiar velocity of the 
main galaxies is smaller than the importance of the  richness of clusters.

The first principal component accounts for about 1/3 of the variance of 
parametres, the second principal component for about 1/4 of the variance. 
However, five principal components are needed to explain more than 90\% of the 
variance of the parametres, thus clusters with their environment are complicated 
objects whose properties cannot be explained with a small number of parametres 
as found also for the dark matter haloes by \citet{2011MNRAS.415L..69J}. 

The locations of clusters in the PC1-PC2 plane shows that unimodal clusters in 
superclusters are located at upper lefthand part of the plot and have larger PC2 
and smaller  (larger negative) PC1 values (for example, clusters 608, 13408, 
25078, and 28508). Rich multimodal clusters of high environmental density value around 
them populate lower and middle righthand area of the biplot (clusters 34276, 
34727, 914, 29587). Multimodal clusters in low environmental density environment 
populate the lefthand lower area of the PC1-PC2 plane (clusters 11474, 11015). 
Unimodal clusters in very rich superclusters  populate the upper righthand area 
of the plane (67116, 63361, and others). On the lefthand area of the plane are 
located isolated multimodal poor clusters (50657, 58323, and others).

\begin{figure}[ht]
\centering
\resizebox{0.45\textwidth}{!}{\includegraphics[angle=0]{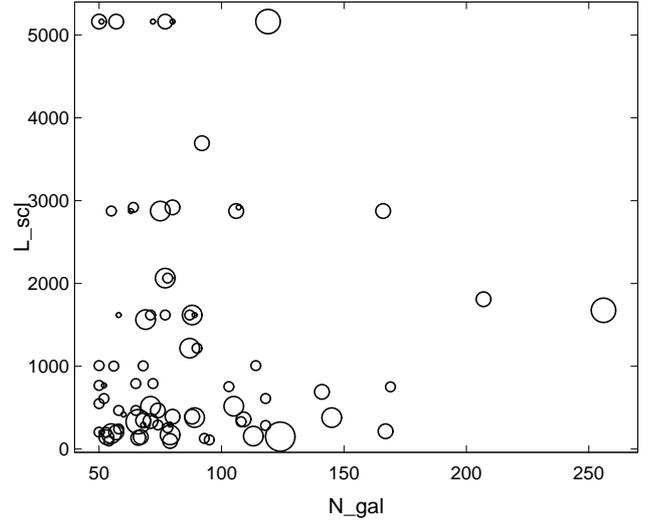}}
\caption{
The number of galaxies in clusters vs. the total luminosity of superclusters
where they reside (in $10^{10}h^{-2} L_{\sun}$). The size of symbols 
is proportional to the number of components in clusters.
}
\label{fig:ngallumscl}
\end{figure}

In Fig.~\ref{fig:ngallumscl} we show for clusters in superclusters the number of 
galaxies in clusters vs. the total luminosity of the host supercluster. Here the 
supercluster of the highest luminosity is SCl~027, the richest system in the 
Sloan Great Wall. This figure shows that this supercluster, as well as other 
superclusters host both multicomponent  and one-component clusters, as a result 
there is no correlation between the host supercluster luminosity and the number 
of components in clusters.

\subsection{Properties of clusters and supercluster morphology}
\label{sect:scl} 

\begin{figure}[ht]
\centering
\resizebox{0.23\textwidth}{!}{\includegraphics[angle=0]{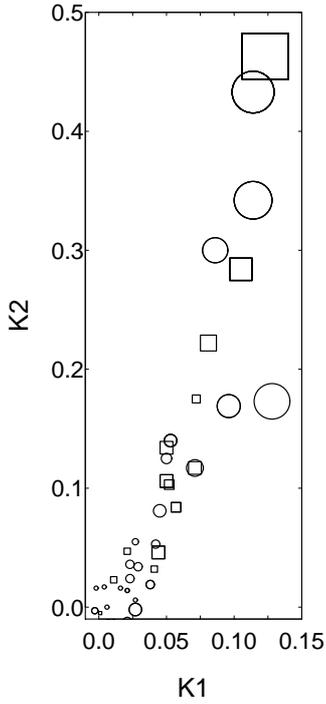}}
\caption{
Shapefinders $K_1$ (planarity) and $K_2$ (filamentarity) for
the superclusters.
The symbol sizes are proportional to the fourth
Minkowski functional $V_3$.
Circles denote the superclusters of spider morphology
and squares denote the superclusters of filament morphology.
}
\label{fig:k12}
\end{figure}

In this section we analyse the properties of clusters in superclusters of 
different morphology. At first we searched for the host superclusters for each 
cluster and found that 80 of our clusters lie in superclusters. Next we 
determined for these superclusters their morphological parametres and types 
using Minkowski functionals and shapefinders, and visual inspection. The 
physical and morphological parametres of superclusters (the values of the fourth 
Minkowski functional (the clumpiness) $V_3$ and the shapefinders $K_1$ (the 
planarity), $K_2$ (the filamentarity), and their ratio (the shape parametre) for 
each supercluster are given in Table~\ref{tab:scl}. According to their overall 
shape superclusters are elongated with the value of the filamentarity $K_2$ 
being larger than the value of the planarity $K_1$. There are only 4 systems 
with the shape parametre $K_1/K_2 > 1.0$ resembling pancakes. The superclusters 
with very small values of the shapefinders have large negative values of the 
shape parametre owing to noisiness in the data. There are 15 superclusters of 
filament morphology, and 35 of spider morphology in our sample. 
Figure~\ref{fig:k12} shows the shapefinders plane for the superclusters with the 
size of symbols proportional to the clumpiness of superclusters takes together 
the morphological information about superclusters. Superclusters with higher 
values of planarity and filamentarity have also larger values of clumpiness and 
therefore more complicated inner morphology. Poor superclusters are mostly of 
spider morphology \citep[we refer for details about the morphological 
information to][]{2011A&A...532A...5E}. Most of them are located close to us, 
they are members of the filaments crossing the void region in front of the Sloan 
Great Wall and other rich superclusters at distances larger than 180~\Mpc.

\begin{figure}[ht]
\centering
\resizebox{0.47\textwidth}{!}{\includegraphics[angle=0]{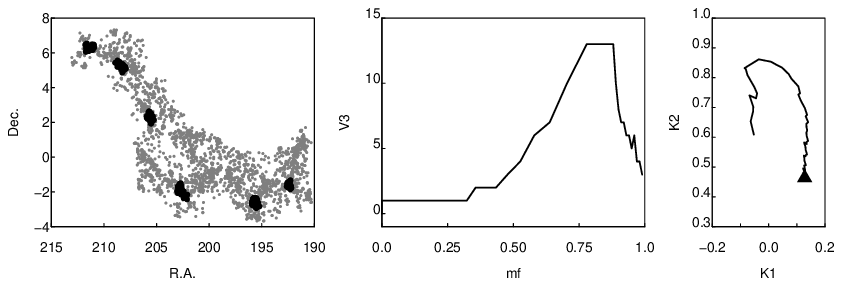}}
\resizebox{0.47\textwidth}{!}{\includegraphics[angle=0]{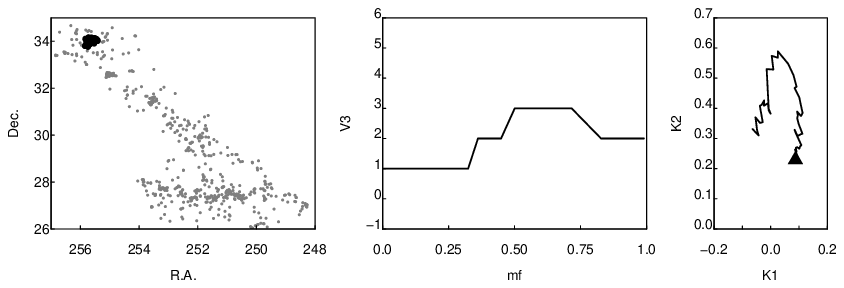}}
\caption{
Sky distribution of galaxies (left panel), the 
fourth Minkowski functional $V_3$ (middle panel) 
and the shapefinder's $K_1$-$K_2$ plane (right panel) for two 
superclusters of filament morphology. Upper row -- the supercluster SCl~027,
lower row -- the supercluster SCl~059. In the left panel black filled circles denote 
galaxies in clusters with at least 50 member galaxies, grey dots denote other
galaxies. On the right panel triangle corresponds to $K_1$ and $K_2$ values at 
the mass fraction $mf = 0$ (the whole supercluster). Mass fraction increases
anti-clockwise along the $K_1$-$K_2$ curve (the morphological signature). 
}
\label{fig:fil}
\end{figure}

\begin{figure}[ht]
\centering
\resizebox{0.47\textwidth}{!}{\includegraphics[angle=0]{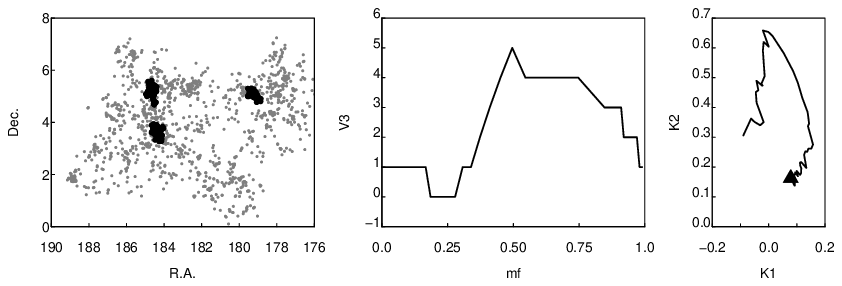}}
\resizebox{0.47\textwidth}{!}{\includegraphics[angle=0]{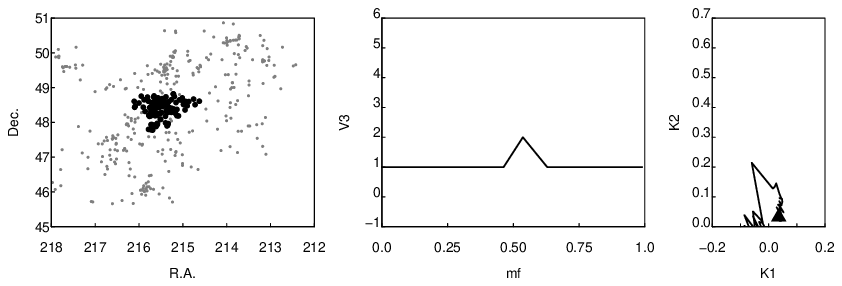}}
\caption{
Sky distribution of galaxies (left panel), the 
fourth Minkowski functional $V_3$ (middle panel) 
and the shapefinder's $K_1$-$K_2$ plane (right panel) for two 
superclusters of spider morphology. Upper row -- the supercluster SCl~019,
lower row -- the supercluster SCl~092. Notations as in Fig.~\ref{fig:fil}.
}
\label{fig:spider}
\end{figure}

In Fig.~\ref{fig:fil} we show the examples of the fourth Minkowski functional 
$V_3$ vs. mass fraction $mf$ and the shapefinders $K_1$ and $K_2$ for two 
superclusters of filament morphology, SCl~027 and SCl~059, in 
Fig.\ref{fig:spider} for superclusters of spider morphology, SCl~019 and SCl~092. 
The superclusters SCl~027 and SCl~019 are the richest two superclusters in the 
Sloan Great Wall \citep{2011ApJ...736...51E, 2011A&A...532A...5E}.  In middle 
panel of these figures we plot the clumpiness $V_3$ vs. the (excluded) mass 
fraction $mf$. At small mass fractions the isodensity surface includes the whole 
supercluster and the value of the 4th Minkowski functional $V_3 = 1$.  As we 
move to higher mass fractions, the iso-surfaces move from the outer supercluster 
boundary towards the higher density parts of a supercluster, and some galaxies 
do not contribute to the supercluster any more. Individual high density regions 
in a supercluster begin to separate from each other, also the holes or tunnels 
may appear, therefore the value of the clumpiness increases. At a certain mass 
fraction $V_3$ has a maximum, showing the largest number of isolated clumps in a 
given supercluster. At still higher mass fraction only the high density peaks 
remain in the supercluster and the value of $V_3$ decreases again. When we 
increase the mass fraction, the changes in the morphological signature accompany 
the changes of the 4th Minkowski functional (right panels of the figures). As 
the mass fraction  increases, at first the planarity $K_1$ almost does not 
change, while the filamentarity $K_2$ increases -- at higher density levels 
superclusters become more filament-like than the whole supercluster. Then also 
the planarity starts to decrease, and at a mass fraction of about $m_f = 0.7$ 
the characteristic morphology of a supercluster changes. We see the crossover 
from the outskirts of a supercluster to the core of a supercluster \citep{e07}. 
The figures of the fourth Minkowski functional and shapefinders for rich 
superclusters with at least 300 member galaxies from SDSS DR7 can be found in 
\citet{2011A&A...532A...5E}.

\begin{table}[ht]
\caption{Properties of clusters in superclusters of spider and filament morphology.}
\begin{tabular}{lcccc} 
\hline\hline 
(1)&(2)& (3)\\      
\hline 
$\mathrm{Parametre}$            &        Spiders &      Filaments \\  
\hline
\\    
$N_{\mathrm{cl}}$      &   33                 &     26               \\ 
$N_{\mathrm{gal}}$     &   77    $\pm$ 15     &  69    $\pm$  15     \\
\\                                               
$N_{\mathrm{comp}}$  &   2.0   $\pm$ 0.46   &  2.0   $\pm$  0.45   \\
$V_{\mathrm{pec}}$     &   291   $\pm$ 60     &  176   $\pm$ 55      \\
$p_{\mathrm{\Delta}}$            &   0.004 $\pm$ 0.013  &  0.004 $\pm$ 0.020   \\
\hline            
                
\label{tab:morf}  
\end{tabular}\\
\tablefoot{ 
Median values and their errors of the number of 3D components, peculiar velocities of main 
galaxies (in $km~s^{-1}$), and $p_{\mathrm{\Delta}}$, and
the numbers of galaxies, $N_{\mathrm{gal}}$
for clusters in superclusters of spider and filament  
morphology (denoted as spiders and filaments)
in a distance interval 180~\Mpc\ $\le D \le $ 300~\Mpc.
}
\end{table}

\begin{table}[ht]
\caption{Properties of clusters in superclusters and isolated clusters.}
\begin{tabular}{lcccc} 
\hline\hline 
(1)&(2)& (3)\\      
\hline 
$\mathrm{Parametre}$            &   Supercluster members & Isolated clusters \\  
\hline
\\    
$N_{\mathrm{cl}}$      &   20                 &     23               \\ 
$N_{\mathrm{gal}}$     &   84    $\pm$ 20     &  58    $\pm$  14     \\
\\                                               
$N_{\mathrm{comp}}$  &   3.0   $\pm$ 0.63   &  3.0   $\pm$  0.59   \\
$V_{\mathrm{pec}}$     &   150   $\pm$ 56     &  201   $\pm$ 59      \\
$p_{\mathrm{\Delta}}$            &   0.001 $\pm$ 0.008  &  0.032 $\pm$ 0.020   \\
\hline            
                
\label{tab:scli}  
\end{tabular}\\
\tablefoot{ 
Median values and their errors of the number of 3D components, peculiar velocities of main 
galaxies (in $km~s^{-1}$), and $p_{\mathrm{\Delta}}$, and
the numbers of galaxies, $N_{\mathrm{gal}}$
for clusters in superclusters and for isolated clusters in a distance interval
120~\Mpc\ $\le D \le $ 180~\Mpc.
}
\end{table}

\begin{figure*}[ht]
\centering
\resizebox{0.9\textwidth}{!}{\includegraphics[angle=0]{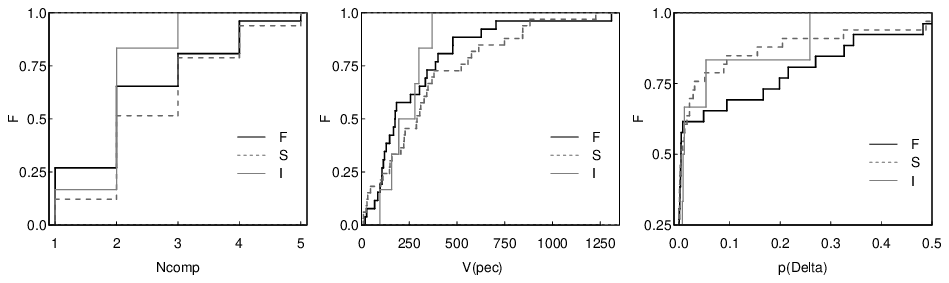}}
\resizebox{0.9\textwidth}{!}{\includegraphics[angle=0]{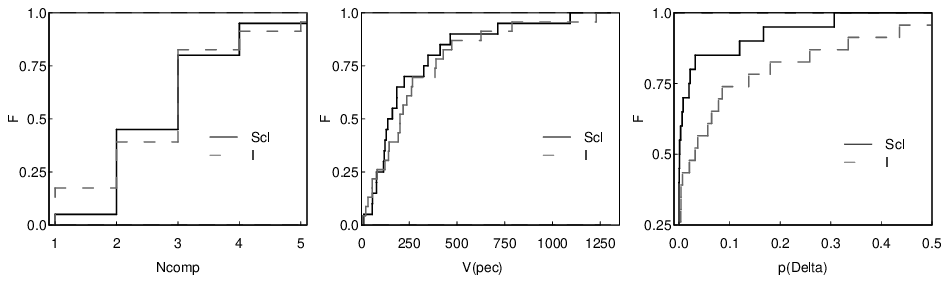}}
\caption{
Cumulative distributions of the numbers of components in clusters, $N_{\mathrm{comp}}$, 
peculiar velocities of cluster main galaxies, $V_{\mathrm{pec}}$  
(in $km~s^{-1}$), and 
p-value of the DS test, $p_{\mathrm{\Delta}}$
for clusters in superclusters of filament morphology (F, black solid line), 
of spider morphology (S, grey dashed line), and for isolated clusters
(I, thin grey solid line) in the distance interval
180~\Mpc\ $\le D \le $ 300~\Mpc\ (upper row), and for clusters in superclusters (Scl, black solid line) and for superclusters
in low density regions (isolated clusters I, grey  solid line)
in the distance interval 120~\Mpc\ $\le D \le $ 180~\Mpc\ (lower row).
}
\label{fig:cummorf}
\end{figure*}

Next we compare the properties of clusters in superclusters of filament and of 
spider morphology, and isolated clusters in two distance intervals. At distances 
up to approximately 180~\Mpc\ most superclusters are poor, and of spider 
morphology. In this distance interval we compare the properties of isolated 
clusters and supercluster members, without dividing them according to the host 
supercluster morphology. At distances of 180~\Mpc\ $\le D \le $ 300~\Mpc\ there 
are only six isolated clusters, therefore we compare the properties of clusters 
in superclusters of different morphology. For a comparison we also show 
parametres of isolated clusters in this distance interval.  We present 
cumulative distributions of the cluster substructure parametres in 
Figure~\ref{fig:cummorf} and the median values of cluster parametres in 
Tables~\ref{tab:morf} and \ref{tab:scli}. 

Table~\ref{tab:morf} and Fig.~\ref{fig:cummorf} (upper row) show that clusters 
in superclusters of spider morphology have higher probabilities to have 
substructure, and larger peculiar velocities of their main galaxies than 
clusters in filament-type superclusters. Clusters  in spider-type superclusters 
are slightly richer than those in filament-type superclusters. Differences in a 
number of components found by 3D normal mixture modelling are small. The 
Kolmogorov-Smirnov test with substructure parametres centred on their means 
showed that in this case the differences between the probalitities to have 
substructure according to the DS test are statistically of very high 
significance ($p < 10^{-6}$), differences between other centred parametres are 
not significant ($p > 0.2$).

Isolated clusters in this distance interval are poor, even poorer than nearby 
isolated clusters, with 52 median number of galaxies. The maximal values of the 
peculiar velocities of their main galaxies, and maximal number of components 
are smaller than those for clusters in superclusters of both types and smaller 
than those for nearby isolated clusters, but only one of them is a one-component 
cluster without significant substructure.

\begin{figure*}[ht]
\centering
\resizebox{0.90\textwidth}{!}{\includegraphics[angle=0]{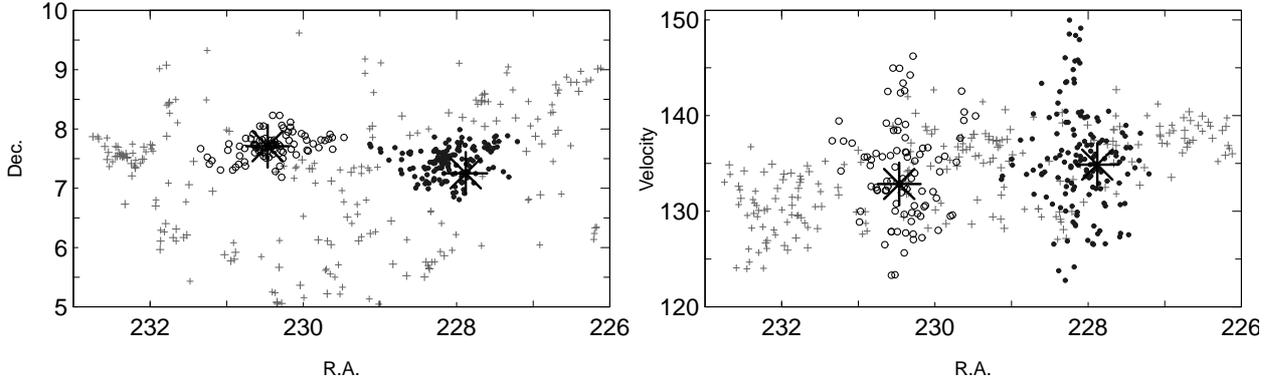}}
\caption{
Distribution of galaxies in R.A. vs. Dec., 
and R.A. vs. velocity (in $10^{2}~km~s^{-1}$) plot
(right panel) in the supercluster SCl~352. Filled circles denote
galaxies in Gr34726, empty circles galaxies in Gr9350. Grey
crosses denote other galaxies in the supercluster. Black stars show the location of 
the main galaxies in both rich cluster.
}
\label{fig:scl352}
\end{figure*}

As an example of a nearby supercluster of spider morphology we show in 
Fig.~\ref{fig:scl352} the distribution of galaxies in SCl~352 in the filament 
crossing the void between nearby superclusters and the Sloan Great Wall 
(Fig.~\ref{fig:richxyz}). This supercluster contain two clusters with at least 
50 member galaxies, Gr9350 and Gr34926. Both have four components and high 
probability to have substructure according to the DS test. The peculiar 
velocities of their main galaxies are small. In the Gr9350 the main galaxy is 
located in the main component of the cluster with  large rms velocities of 
galaxies (the finger-of-god, seen in the right panel of the figure). In Gr34926 
the main galaxy is located in another component with smaller rms velocities of 
galaxies. We see in this figure rich inner structure of the supercluster, where 
clusters and groups of galaxies are connected by galaxy filaments. Some 
components of rich clusters may be infalling, a hint that clusters are not 
virialised yet.

\begin{figure*}[ht]
\centering
\resizebox{0.90\textwidth}{!}{\includegraphics[angle=0]{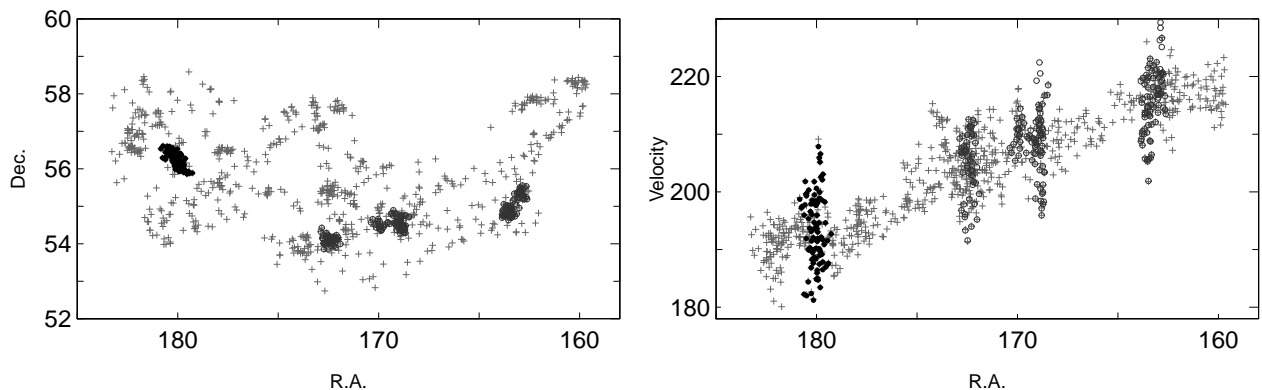}}
\caption{
Distribution of galaxies in R.A. vs. Dec., 
and R.A. vs. velocity (in $10^{2}~km~s^{-1}$) plot
(right panel) in the supercluster SCl~211 (the Ursa 
Major supercluster). 
Filled circles denote
galaxies in Gr5217 (Abell cluster A 1436), empty circles galaxies in 
other clusters with at least 50
member galaxies. Grey crosses denote other galaxies in the supercluster. 
}
\label{fig:scl211}
\end{figure*}

Nearby clusters in superclusters are richer than more distant clusters in 
superclusters. They have larger probabilities to have substructure than clusters 
in more distant superclusters, but smaller peculiar velocities of their main 
galaxies. We use flux-limited sample of galaxies to define clusters, and nearby 
clusters contain galaxies of lower luminosity than distant clusters. It is 
possible that these clusters have substructure in their outer regions formed by 
fainter galaxies, absent in more distant sample, and this may explain the 
difference between them. They also have different global environment: nearby 
superclusters are located in poor filaments surrounded by voids. More distant 
superclusters are richer and form several chains of rich superclusters 
\citep[for details about the environmental environment of superclusters we refer 
to][]{2011A&A...532A...5E}. Next we plan to analyse the galaxy content of 
clusters in more detail, and also a larger sample of clusters and superclusters, 
to understand better the difference between clusters. 

Isolated clusters are poorer than clusters in superclusters. The number of 
components in isolated clusters is close to that for supercluster members in the 
same distance interval. The peculiar velocities of the main galaxies in isolated 
clusters are larger than in supercluster members, but maximal values of the 
peculiar velocities of main galaxies in isolated clusters are smaller than those 
in supercluster members. The DS test shows that the probability to have 
substructure is larger among nearby supercluster member clusters than in 
isolated clusters. Among six distant isolated clusters only one has no
significant substructure. 

\subsection{Examples of selected clusters}
\label{sect:selcl} 

{\it Most luminous clusters}. Our sample contains seven clusters of very high 
luminosity (Table~\ref{tab:lumcl}). All of them have been identified with Abell 
clusters. They are located in high density cores of superclusters 
(Sect.\ref{sect:lss}. Five of seven most luminous clusters are located in 
superclusters of spider morphology. Among them are the cluster Gr34727 and the 
cluster Gr39489 in superclusters of the dominant supercluster plane (SCl~007, 
and SCl~099, the Corona Borealis supercluster, correspondingly). The Corona 
Borealis supercluster and clusters in it have been studied by 
\citet{1998ApJ...492...45S} and \citet{2009MNRAS.396...53P}. One of the high-
luminosity clusters in the filament-type supercluster is Gr914 (Abell cluster 
\object{A 1750}) in the richest supercluster of the Sloan Great Wall, SCl~027, 
another -- Gr29587 (A2142) in SCl~001 in the dominant supercluster plane. The 
luminosity density around A2142 in SCl~001 is very high \citep[for details and 
references see][]{2011A&A...532A...5E}. The clusters   A1750, A2028, A2029, 
A2065, A2069,  and A2142 are  (probably merging) X-ray clusters 
\citep{2000ApJ...541..542M, belsole04, 2004ApJ...616..178C, 
2008A&A...479..307B, 2009ApJ...704.1349O, 2010A&A...522A..34G}. All high-
luminosity clusters have multiple components, high probability to have 
substructure, and most of them have large peculiar velocities of their main 
galaxies.

{\it Most unimodal and most multimodal clusters}. 
Some of our clusters are 
unimodal according to all the tests applied in E12, they are one-component 
systems with very low probability of substructure, the sky distribution of their 
member galaxies is symmetrical, and the galaxy velocity distribution is 
Gaussian. We list them in Table~\ref{tab:unicl}. There are also multimodal 
clusters according to all the tests with multiple components, asymmetrical 
galaxy distribution and non-Gaussian distribution of galaxy velocities 
(Table~\ref{tab:multicl}. We marked them in luminosity-density plots in 
Fig.~\ref{fig:lumden}. Tables~\ref{tab:unicl} and ~\ref{tab:multicl} show that 
unimodal clusters are, in average, poorer and of lower luminosity than 
multimodal clusters. Figure~\ref{fig:d4d} showed that unimodal clusters are 
typically located in a lower density environment, but in Fig.~\ref{fig:lumden} 
we saw that some of them reside also in high-density cores of rich 
superclusters.  Of seven most unimodal clusters four are located in 
superclusters of filament morphology, two in spider-type superclusters and one 
is isolated. In contrary, of seven most multimodal clusters five are located in 
spider-type superclusers, and only two in filament-type superclusters. There are 
two superclusters in our sample which both hosts one most multimodal and one 
most unimodal cluster - SCl~027, and SCl~211. In Fig.~\ref{fig:scl211} we plot 
the distribution of galaxies in the supercluster SCl~211 (the Ursa Major 
supercluster). Here Gr5217  is located approximately at $R.A. = 180$ and $Dec. = 
56.2$ degrees, and Gr28387 at $R.A. = 170$ and $Dec. = 54.5$ degrees. 

Unimodal cluster Gr25078 (Abell cluster A1650) in the core of SCl~027 is a 
compact X-ray source, possibly located at a cold spot in the CMB \citep{udo04}. 
The cluster  A1809 in the supercluster SCl~027  is also a X-ray cluster. 
Multimodal clusters A1291, A1983 and A671 are also X-ray clusters 
\citep{2003A&A...408....1P, 2011A&A...532A...5E, 2012MNRAS.tmp.2432P}.

\section{Discussion and conclusions}
\label{sect:discussion} 

We studied the environment of rich clusters from SDSS DR8, defined by the 
environmental luminosity density around clusters, and by membership of clusters 
in superclusters of different morphology. We found a correlation between the 
environmental density around clusters and the presence of substructure in 
clusters. However, both multimodal and unimodal clusters can be found in regions 
of low and high environmental density, and correlations with the environmental 
density are not strong. In the study of the substructure of the richest clusters 
of the Sloan Great Wall \citet{2010A&A...522A..92E} found clusters with 
substructure in both rich and poor superclusters of the Wall. In this study we 
showed using a larger sample of superclusters and clusters that superclusters of 
different richness may host both multicomponent  and one-component clusters, and 
there is no correlation between the host supercluster richness (luminosity) and 
the multimodality of clusters. In higher density environment the peculiar 
velocities of the main galaxies in clusters are larger, a suggestion that by 
this measure also clusters in our sample are dynamically more active in high 
density environments. Isolated clusters are poorer and have smaller maximal 
number of components than cluster in superclusters in the same distance 
interval. \citet{2002MNRAS.329L..47P}, \citet{2004ogci.conf...19P}, 
\citet{2007MNRAS.377.1785R} and \citet{2007ApJ...666L...5E} showed that 
dynamically younger clusters with more substructure are more strongly clustered 
than overall cluster population. They used  cluster centroid shift and DS test 
results as indicators of the dynamical state of clusters.  We only use the data 
about rich clusters in our sample, and majority of them (80\%) show 
multimodality according to at least one test in E12 while other studies included 
data also about poorer systems. This is probably the reason why we found weaker 
correlations between the presence of substructure and cluster environment than 
in other studies. 

We also found clusters in our sample with  almost no close galaxies or galaxy 
filaments (within the SDSS survey magnitude limits) in the sky distribution. One 
example of such a cluster is given in Fig.~\ref{fig:scl211}. This is the 
unimodal cluster Gr5217 (A1436), surrounded by an almost empty region without 
visible galaxies, seen also in Figure~1b by \citet{2007AstL...33..211K}. Other 
rich clusters in this system are connected by galaxy filaments. This contradicts 
with understanding that rich clusters are located at the intersections of galaxy 
filaments. One possible reason for that may be that all brighter galaxies in 
this region have already merged into the cluster. \citet{2009AstBu..64....1K} 
write that A1436 is probably relaxing after a recent merger, which agrees with 
our interpretation. We found about ten of such clusters in our sample, both one-
component and multicomponent ones, and continue to study them to better 
understand the relation between rich clusters and their small and large scale 
environment.

Cosmological simulations of the future evolution of the structure in the 
Universe predict that future superclusters (at $a = 100$, $a$ is the expansion 
factor) are much more spherical than present-day superclusters. Clusters in 
superclusters will merge into a single cluster in the far future. In other 
words, the differences noted in this study between superclusters of different 
morphology, and clusters in them, may disappear in the distant future \citep[see 
][for a review and references]{2009MNRAS.399...97A}.  

In the framework of the hierarchical formation of the structure galaxy clusters 
are located at the intersections of filaments which form the cosmic web. Matter 
flows through filaments from lower density regions into clusters. The merging 
and growth of dark matter haloes have been studied with cosmological simulations 
\citep[][and references therein]{1992ApJ...393..477R, 2002MNRAS.336..112M, 
2008MNRAS.388.1537M, 2010MNRAS.406.2267F} which show that the late time 
formation of the main haloes and the number of recent major mergers can cause 
the late time subgrouping of haloes and the presence of substructure 
\citep{2008ApJ...682L..73S, 2010A&A...522A..92E, 2012MNRAS.419.1576P}. In high-
density regions groups and clusters of galaxies form early and could be more 
evolved dynamically \citep{tempel09}, but in high-density regions  the 
velocities of haloes in the vicinity of larger haloes are high 
\citep{2005A&A...436...17E} and the possibility of mergers is also high. As a 
result in high-density regions clusters have a larger amount of substructure and 
higher peculiar velocities of their main galaxies than in low-density regions. 

Superclusters of spider morphology have richer inner structure than 
superclusters of filament morphology with large number of filaments between 
clusters in them.  This may lead to the differences noted in this study: 
clusters in superclusters of spider morphology are dynamically younger than 
clusters in superclusters of filament morphology with their higher probability 
to have substructure and larger peculiar velocities of main galaxies. Five of 
seven most multimodal clusters are located in superclusters of spider 
morphology, while four of six most unimodal clusters in superclusters reside in 
filament-type superclusters.

One example of the high-luminosity multimodal cluster in the spider-type supercluster
is Gr34727 (Abell cluster A2028).  \citet{2010A&A...522A..34G}
suggest that the subclusters of A 2028 are probably merging to produce
a more massive cluster. Merging X-ray clusters have been found also in the high-density
cores of other superclusters \citep{rose02, 2000MNRAS.312..540B}.

\citet{2011ApJ...736...51E} calculated the peculiar velocities of the main 
galaxies in groups from the two richest superclusters in the Sloan Great Wall 
and found that groups in the supercluster SCl~027  of filamentary morphology 
(see Table~\ref{tab:scl}) are dynamically more active with their larger peculiar 
velocities of main galaxies than groups in SCl~019 of spider morphology. In 
SCl~027 the environmental densities are very high, this may be the reason why in 
this supercluster of filament morphology clusters show higher dynamical activity 
than another supercluster of the Sloan Great Wall.

\citet{2011A&A...531A..75E, 2011A&A...534A.128E}, and 
\citet{2011A&A...531A.149S} showed how the syncronisation of phases of density 
waves of different scales affect the richness of galaxy systems: the larger the 
scale of density waves, where the maxima of waves of different scales have close 
positions, the larger are the masses of galaxy clusters. In lower density 
regions the formation of rich clusters is suppressed by the combined negative 
sections of medium- and large- scale density perturbations. This process makes 
voids empty of galaxies and their systems, clusters of galaxies can be found in 
filaments crossing the voids, and they are not so rich as galaxy clusters in 
higher density regions. This is what we found in this study. Richer clusters 
from our sample are located in regions of high environmental density, in 
agreement also with earlier results about observations and simulations 
\citep[][and references therein]{e2003a, e2003b, 2005A&A...436...17E, 
2011arXiv1109.4169C}, where positive sections of medium- and large-scale density 
perturbations combine. The most luminous clusters are located in high-density 
cores of rich superclusters, five of seven most luminous clusters are in spider-
type superclusters. From the other hand, \citet{2011A&A...532A...5E} studied the 
morphology of superclusters from SDSS DR7 and found that among them there are no 
compact and very luminous superclusters. Poor superclusters have lower 
luminosities and they host clusters of lower luminosity. 

Correlations between the cluster parametres and the environmental density are 
stronger at small smoothing lengths and become weaker as we increase the 
smoothing length, in agreement with \citet{e2003a} and \citet{e2003a} who showed 
that the properties of galaxy groups depend on environmental density up to 
scales of about 15 -- 20 \Mpc. Simulations show that also halo spin, overall 
shape and other properties depend on environment \citep[][and references 
therein]{2005A&A...436...17E, 2007MNRAS.375..489H, 2010MNRAS.408.1818W, 
2011MNRAS.413..301N, 2011MNRAS.413.1973W, 2012arXiv1201.5794C} although 
\citet{2011MNRAS.415L..69J} and \citet{2011MNRAS.416.2388S} showed that 
dependence on environment is weaker than found in other studies.

Summarising, our conclusions are as follows.

\begin{itemize}
\item[1)]
Clusters from our sample are located in density peaks in filaments
crossing voids and in superclusters.
\item[2)]
The values of the environmental densities around multimodal clusters (i.e. those with large
number of components,  
high probabilities to have substructure, and
large peculiar velocities of their main galaxies) are higher than
those around unimodal clusters.
\item[3)] 
We determined
the values of the fourth Minkowski functional and shapefinders,
and morphological types for each supercluster hosting clusters from our sample.
Of 50 superclusters hosting rich clusters 35 are of spider type and 15 of
filament type. 
\item[4)] 
Clusters in superclusters of spider morphology have 
 higher probabilities to have substructure, and higher
values of the peculiar velocities of their main galaxies than
clusters in superclusters of filament morphology.
\item[5)]
High-luminosity clusters reside in the cores of rich superclusters.
Five out of seven high-luminosity clusters belong to superclusters of spider 
morphology. The most multimodal clusters are preferentially located in 
spider-type superclusters, while four of seven most unimodal clusters reside
in filament-type superclusters.
\item[6)]
Isolated clusters are poorer and they 
have smaller maximal number of components and lower maximal (but higher median) 
values of the  peculiar velocities of their main galaxies
than supercluster members. 
\item[7)]  
High luminosity superclusters have higher values of the environmental densities
and peak densities than low luminosity superclusters.
\end{itemize}

Our study shows the importance of the role of superclusters as high density 
environment which affects  the properties of galaxies and galaxy systems in them 
\citep{2004ogci.conf...19P, 2005A&A...443..435W, 2006MNRAS.371...55H, e07, 
2008MNRAS.388.1152P, tempel09, 2010ASPC..423...81F, 2011A&A...529A..53T, 
2011ApJ...736...51E}. Earlier studies of galaxy superclusters revealed that 
while according to their overall properties superclusters can be described with 
a small number of parametres \citep{2011A&A...535A..36E}, the analysis of the 
morphology and galaxy and group content of the richest superclusters from  the 
2dF Galaxy Redshift Survey and from the Sloan Great Wall with SDSS data 
\citep{2008ApJ...685...83E, 2011ApJ...736...51E} demonstrate a large variety of 
supercluster morphologies and differences in the distribution of galaxies from 
various populations, and groups of galaxies in superclusters. Observations have 
already revealed large differences between galaxy and group content in high-
redshift superclusters \citep{2009AJ....137.4867L, 2011A&A...532A..57S}. Such a 
large variety of observational properties is not yet recovered by simulations 
\citep[][and references therein]{e07, 2011ApJ...736...51E} and not well 
understood from observations. As a next step we plan to study the properties of 
a large sample of groups and clusters in superclusters of different morphology 
to better understand the differences and similarities between them.

\section*{Acknowledgments}

We thank our referee for  very detailed comments which helped to improve the paper. 

We are pleased to thank the SDSS Team for the publicly available data
releases.  
Funding for the Sloan Digital Sky Survey (SDSS) and SDSS-II has been
  provided by the Alfred P. Sloan Foundation, the Participating Institutions,
  the National Science Foundation, the U.S.  Department of Energy, the
  National Aeronautics and Space Administration, the Japanese Monbukagakusho,
  and the Max Planck Society, and the Higher Education Funding Council for
  England.  The SDSS Web site is \texttt{http://www.sdss.org/}.
  The SDSS is managed by the Astrophysical Research Consortium (ARC) for the
  Participating Institutions.  The Participating Institutions are the American
  Museum of Natural History, Astrophysical Institute Potsdam, University of
  Basel, University of Cambridge, Case Western Reserve University, The
  University of Chicago, Drexel University, Fermilab, the Institute for
  Advanced Study, the Japan Participation Group, The Johns Hopkins University,
  the Joint Institute for Nuclear Astrophysics, the Kavli Institute for
  Particle Astrophysics and Cosmology, the Korean Scientist Group, the Chinese
  Academy of Sciences (LAMOST), Los Alamos National Laboratory, the
  Max-Planck-Institute for Astronomy (MPIA), the Max-Planck-Institute for
  Astrophysics (MPA), New Mexico State University, Ohio State University,
  University of Pittsburgh, University of Portsmouth, Princeton University,
  the United States Naval Observatory, and the University of Washington.

The present study was supported by the Estonian Science Foundation
grants No. 8005, 7765, 9428, and MJD272, by the Estonian Ministry for Education and
Science research project SF0060067s08, and by the European Structural Funds
grant for the Centre of Excellence "Dark Matter in (Astro)particle Physics and
Cosmology" TK120. This work has also been supported by
ICRAnet through a professorship for Jaan Einasto.
P.N. was supported by the Academy of Finland, P.H. by Turku University Foundation. 
V. M. was supported by the Spanish MICINN CONSOLIDER projects 
ATA2006-14056 and CSD2007-00060, including FEDER contributions, 
and by the Generalitat Valenciana project of excellence PROMETEO/2009/064.

\bibliographystyle{aa}
\bibliography{env.bib}

\begin{appendix}

\section{Multimodality of clusters}
\label{sect:multi}

We employ two 3D methods to search for substructure in clusters. With {\it 
Mclust} package \citep{fraley2006} from {\it R}, an open-source free statistical 
environment developed under the GNU GPL \citep[][\texttt{http://www.r-
project.org}]{ig96} we search for an optimal model for the clustering of the 
data among models with varying shape, orientation and volume, under assumption 
that the multivariate sample is a random sample from a mixture of multivariate 
normal distributions. {\it Mclust} finds the optimal number of components
and calculates for every galaxy the probabilities 
to belong to any of the components which are used to calculate the uncertainties 
for galaxies to belong to a component. The mean uncertainty for the full sample 
is used  as a statistical estimate of the reliability of the results. The 
calculations in E12 showed that uncertainties are small, their values can be 
found in online tables of E12.
We tested how the possible errors in the line-of-sight positions of galaxies 
affect the results  of {\it Mclust}, shifting randomly the peculiar velocities 
of galaxies 1000 times and searching  each time for the components with {\it 
Mclust}. The random shifts were chosen from a Gaussian
distribution with the dispersion equal to the sample velocity dispersion of galaxies 
in a cluster.  The number of the components found by {\it Mclust} remained unchanged, 
demonstrating that the results of {\it Mclust} are not sensitive to such errors.

The Dressler-Shectman (DS) test  
searches for deviations of the 
local velocity mean and dispersion from the cluster mean values. The algorithm 
starts by calculating the mean velocity ($v_\mathrm{local}$) and the velocity dispersion 
($\sigma_\mathrm{local}$) for each galaxy of the cluster, using its $n$ nearest 
neighbours. These values of local kinematics are compared with the mean velocity 
($v_c$) and the velocity dispersion ($\sigma_c$) determined for the entire cluster of $N_{gal}$ 
galaxies. The differences between the local and global kinematics are 
quantified by 

$$ \delta_i^2 = (n+1)/\sigma_c^2[(v_\mathrm{local}-v_c)^2 + (\sigma_\mathrm{local}-\sigma_c)^2]. $$

The cumulative deviation $\Delta = \Sigma \delta_i$ is  used as a statistic for 
quantifying (the significance of) the substructure. The results of  the DS-test 
depend on the number of local galaxies $n$. We use $n = \sqrt N_{gal}$, as suggested 
by \citet{1996ApJS..104....1P}. The $\Delta$ statistic for each cluster should 
be calibrated by Monte Carlo simulations. In Monte Carlo models the velocities 
of galaxies are randomly shuffled among the positions. We ran 25000 models for 
each cluster and calculated every time $\Delta_\mathrm{sim}$. The significance 
of having substructure (the  $p_\Delta$-value) can be quantified by the ratio 
$N(\Delta_\mathrm{sim} > \Delta_{obs})/N_\mathrm{sim}$ --  the ratio of the 
number of simulations in which the value of $\Delta$ is larger than the observed 
value, and the total number of simulations. The smaller the 
$p_\Delta$-value, the 
larger is the probability of substructure.

\section{Luminosity density field and superclusters}
\label{sect:DF}

To calculate the luminosity density field, we  calculate
the luminosities of groups first. In flux-limited samples, galaxies outside
the observational window remain unobserved. To take into account
the luminosities of the galaxies that lie outside the sample limits also
 we multiply the observed galaxy
luminosities by the weight $W_d$.  The distance-dependent weight
factor $W_d$ was calculated as  
\begin{equation}
    W_d =  {\frac{\int_0^\infty L\,n
    (L)\mathrm{d}L}{\int_{L_1}^{L_2} L\,n(L)\mathrm{d}L}} ,
    \label{eq:weight}
\end{equation}
where $L_{1,2}=L_{\sun} 10^{0.4(M_{\sun}-M_{1,2})}$ are the luminosity 
limits of the observational window at a distance $d$, corresponding to the 
absolute magnitude limits of the window $M_1$ and $M_2$; we took 
$M_{\sun}=4.64$\,mag in the $r$-band \citep{2007AJ....133..734B}. 
Due to their peculiar velocities, 
the distances of galaxies are somewhat uncertain; if the galaxy belongs to a 
group, we use the group distance to determine the weight factor. 
Details of the calculations of weights are given also in
 \citet{2011A&A...529A..53T} and in \citet{2011A&A...535A..36E}.

To calculate a luminosity density field, 
we convert the spatial positions of galaxies $\mathbf{r}_i$ 
and their luminosities  $L_i$ into
spatial (luminosity) densities using kernel densities
\citep{silverman86}:
\begin{equation}
    \rho(\mathbf{r}) = \sum_i K\left( \mathbf{r} - \mathbf{r}_i; a\right) L_i,
\end{equation}
where the sum is over all galaxies, and $K\left(\mathbf{r};
a\right)$ is a kernel function of a width $a$. Good kernels
for calculating densities on a spatial grid are generated by box splines
$B_J$. Box splines are local and they are interpolating on a grid:
\begin{equation}
    \sum_i B_J \left(x-i \right) = 1,
\end{equation}
for any $x$ and a small number of indices that give non-zero values for $B_J(x)$.
We use the popular $B_3$ spline function:
        \begin{eqnarray}
        B_3(x)&=&\left(|x-2|^3-4|x-1|^3+6|x|^3-\right.\nonumber\\
                &&\left.-4|x+1|^3+|x+2|^3\right)/12.
        \end{eqnarray}
The (one-dimensional) $B_3$ box spline kernel $K_B^{(1)}$ of the width $a$ is defined as
\begin{equation}
    K_B^{(1)}(x;a,\delta) = B_3(x/a)(\delta / a),
\end{equation}
where $\delta$ is the grid step. This kernel differs from zero only
in the interval $x\in[-2a,2a]$. It is close to a Gaussian with $\sigma=0.6$ in the
region $x\in[-a,a]$, so its effective width is $2a$ \citep[see, e.g.,][]{saar09}.
The kernel preserves the
interpolation property exactly for all values of $a$ and $\delta$,
where the ratio $a/\delta$ is an integer. (This kernel can be used also if this ratio
is not an integer, 
and $a \gg \delta$; the kernel sums to 1 in this case, too, with a very small error.)
This means that if we apply this kernel to $N$ points on a one-dimensional grid,
the sum of the densities over the grid is exactly $N$.
 
The three-dimensional kernel $K_B^{(3)}$
is given by the direct product of three one-dimensional kernels:
\begin{equation}
    K_B^{(3)}(\mathbf{r};a,\delta) \equiv K_3^{(1)}(x;a,\delta) K_3^{(1)}(y;a,\delta) K_3^{(1)}(z;a,\delta),
\end{equation}
where $\mathbf{r} \equiv \{x,y,z\}$. Although this is a direct product,
it is isotropic to a good degree \citep{saar09}.

The densities were calculated on a cartesian grid based on the SDSS $\eta$, 
$\lambda$ coordinate system. The grid coordinates are calculated according to 
Eq.\ref{eq:xyz}. We used an 1~\Mpc\ step grid and chose the kernel width 
$a=8$~\Mpc. This kernel differs from zero within the radius 16~\Mpc, but 
significantly so only inside the 8~\Mpc\ radius. As a lower limit for the volume 
of superclusters we used the value $(a/2)$~\Mpc$^3$ (64 grid cells). We also 
used density field with the kernel widths $a=4$~\Mpc,  $a=8$~\Mpc, and 
$a=16$~\Mpc\ to characterise the environmental density around clusters. Before 
extracting superclusters we apply the DR7 mask constructed by P.~Arnalte-Mur 
\citep{martinez09, 2010arXiv1012.1989J} to the density field and convert 
densities into units of mean density. The mean density is defined as the average 
over all pixel values inside the mask. The mask is designed to follow the edges 
of the survey and the galaxy distribution inside the mask is assumed to be 
homogeneous.

\section{Minkowski functionals and shapefinders} 
\label{sect:MF}

For a given surface the four Minkowski functionals (from the first to the
fourth) are proportional to the enclosed volume $V$, the area of the surface
$S$, the integrated mean curvature $C$, and the integrated Gaussian curvature
$\chi$. 
Consider an
excursion set $F_{\phi_0}$ of a field $\phi(\mathbf{x})$ (the set
of all points where the density is higher than a given limit,
$\phi(\mathbf{x}\ge\phi_0$)). Then, the first
Minkowski functional (the volume functional) is the volume of 
this region (the excursion set):
\begin{equation}
\label{mf0}
V_0(\phi_0)=\int_{F_{\phi_0}}\mathrm{d}^3x\;.
\end{equation}
The second Minkowski functional is proportional to the surface area
of the boundary $\delta F_\phi$ of the excursion set:
\begin{equation}
\label{mf1}
V_1(\phi_0)=\frac16\int_{\delta F_{\phi_0}}\mathrm{d}S(\mathbf{x})\;,
\end{equation}
(but it is not the area itself, notice the constant).
The third Minkowski functional is proportional to the
integrated mean curvature $C$ of the boundary:
\begin{equation}
\label{mf2}
V_2(\phi_0)=\frac1{6\pi}\int_{\delta F_{\phi_0}}
    \left(\frac1{R_1(\mathbf{x})}+\frac1{R_2(\mathbf{x})}\right)\mathrm{d}S(\mathbf{x})\;,
\end{equation}
where $R_1(\mathbf{x})$ and $R_2(\mathbf{x})$ 
are the principal radii of curvature of the boundary.
The fourth Minkowski functional is proportional to the integrated
Gaussian curvature (the Euler characteristic) 
of the boundary:
\begin{equation}
\label{mf3}
V_3(\phi_0)=\frac1{4\pi}\int_{\delta F_{\phi_0}}
    \frac1{R_1(\mathbf{x})R_2(\mathbf{x})}\mathrm{d}S(\mathbf{x})\;.
\end{equation}
At high (low) densities this functional gives us the number of isolated 
clumps (void bubbles) in the sample 
\citep{mar03,saar06}:

\begin{equation}
\label{v3}
V_3=N_{\mbox{clumps}} + N_{\mbox{bubbles}} - N_{\mbox{tunnels}}.
\end{equation}

As the argument labelling the isodensity surfaces, we chose the (excluded) mass
fraction $mf$ -- the ratio of the mass in the regions with the density {\em lower}
than the density at the surface, to the total mass of the supercluster. When
this ratio runs from 0 to 1, the iso-surfaces move from the outer limiting
boundary into the centre of the supercluster, i.e., the fraction $mf=0$
corresponds to the whole supercluster, and $mf=1$ -- to its highest density
peak.

The first three functionals were used to calculate the shapefinders
\citep{sah98,sss04,saar09}.  The shapefinders are defined as a set of
combinations of Minkowski functionals: $H_1=3V/S$ (thickness),
$H_2=S/C$ (width), and $H_3=C/4\pi$ (length).  The shapefinders have
dimensions of length and are normalized to give $H_i=R$ for a sphere
of radius $R$.  For smooth (ellipsoidal) surfaces, the shapefinders $H_i$
follow the inequalities $H_1\leq H_2\leq H_3$.  Oblate ellipsoids (pancakes)
are characterised by $H_1 << H_2 \approx H_3$, while prolate ellipsoids
(filaments) are described by $H_1 \approx H_2 << H_3$.
\citet{sah98} also defined  two dimensionless
shapefinders $K_1$ (planarity) and $K_2$ (filamentarity): 
$K_1 = (H_2 - H_1)/(H_2 + H_1)$ and $K_2 = (H_3 -
H_2)/(H_3 + H_2)$.
In the $(K_1,K_2)$-plane filaments are located near the $K_2$-axis,
pancakes near the $K_1$-axis, and ribbons along the diameter, connecting 
the spheres at the origin with the ideal ribbon at $(1,1)$. 
In \citet{e07} we calculated typical morphological signatures 
of a series of empirical models that serve as  morphological 
templates to compare with the characteristic curves for superclusters
in the $(K_1,K_2)$-plane.

\section{Data on selected clusters} 
\label{sect:speccl}

\begin{table*}[ht]
\caption{Data on most luminous clusters}
\begin{tabular}{rrrrrrrrrrrrrr} 
\hline\hline  
(1)&(2)&(3)&(4)&(5)& (6)&(7)&(8)&(9)& (10) & (11)&(12)& (13)& (14)\\      
\hline 
 ID   &$N_{\mathrm{gal}}$ & $L_{\mathrm{tot}}$ & $\sigma$ & $r_{\mathrm{vir}}$ &
 $V_{\mathrm{pec}}$ & $N_{\mathrm{comp}}$ & $p_{\mathrm{\Delta}}$& $D4$ & $D8$ & $D16$ & $ID_{\mathrm{scl}}$ & Type & Abell ID \\
 & & $10^{10} h^{-2} L_{\sun}$&$km~s^{-1}$&$h^{-1}$ Mpc& $km~s^{-1}$ &&&&&&& \\
\hline
  914 & 119 &  227 &  657 & 0.83 & -704 & 5 & $< 10^{-4}$ & 49.5 & 13.2 & 4.0 & 27 & F & \object{A 1750} \\
29587 & 207 &  365 &  740 & 0.87 &  334 & 3 & $< 10^{-4}$ & 79.3 & 21.3 & 5.8 &  1 & F & \object{A 2142}   \\
34727 & 256 &  351 &  825 & 1.25 & -290 & 5 & $< 10^{-4}$ & 51.3 & 16.1 & 5.1 &  7 & S & \object{A 2028},\object{A 2029},\object{A 2033},\object{A 2040}   \\
38087 & 169 &  209 &  541 & 0.84 &  844 & 2 & 0.031 &           51.1 & 13.9 & 3.7 & 24 & S & \object{A 1173},\object{A 1187},\object{A 1190},\object{A 1203}  \\
39489 & 166 &  188 & 1061 & 0.72 & -521 & 3 & $< 10^{-4}$ & 43.5 & 10.7 & 4.0 & 99 & S & \object{A 2056},\object{A 2065}   \\
68625 &  92 &  301 &  874 & 0.79 &  825 & 3 & $< 10^{-4}$ & 72.6 & 20.0 & 6.3 &  3 & S & \object{A 2069}  \\
73088 & 141 &  184 &  631 & 0.71 & -224 & 3 & $< 10^{-4}$ & 47.7 & 11.5 & 3.2 & 92 & S & \object{A 1904}  \\
\label{tab:lumcl}  
\end{tabular}\\
\tablefoot{                                                                                 
Columns are as follows:
1: ID of the cluster;
2: the number of galaxies in the cluster, $N_{\mathrm{gal}}$;
3: total luminosity of  the cluster;
4: rms velocity of the cluster;
5: virial radius of the cluster;
6:  peculiar velocity of the main galaxy;
7: the number of components in the cluster, $N_{\mathrm{comp}}$;
8: p-value of  the DS test;
9--11: environmental density around the cluster, at smoothing lengths 4, 8, and 16 \Mpc\
(in units of the mean density);
12: ID of the supercluster where the cluster resides. 
13: morphological type of the supercluster 
14: Abell ID of the cluster
}
\end{table*}

\begin{table*}[ht]
\caption{Data on unimodal clusters}
\begin{tabular}{rrrrrrrrrrrrrr} 
\hline\hline  
(1)&(2)&(3)&(4)&(5)& (6)&(7)&(8)&(9)& (10) & (11)&(12) & (13)&(14)\\      
\hline 
 ID   &$N_{\mathrm{gal}}$ & $L_{\mathrm{tot}}$ & $\sigma$ & $r_{\mathrm{vir}}$ &
 $V_{\mathrm{pec}}$ & $N_{\mathrm{comp}}$ & $p_{\mathrm{\Delta}}$& $D4$ & $D8$ & $D16$ & $ID_{\mathrm{scl}}$&Type & Abell ID\\
 & & $10^{10} h^{-2} L_{\sun}$&$km~s^{-1}$&$h^{-1}$ Mpc& $km~s^{-1}$ &&&&&&& \\
\hline
  608 &  60 &  132 &  532 & 0.66 & -228 & 1 & 0.884 &  33.4 &   7.6 &  2.7 &  569 & S & \object{A 2175}  \\
 5217 &  89 &   94 &  577 & 0.61 &  -19 & 1 & 0.216 &  27.3 &   8.0 &  3.2 &  211 & F & \object{A 1436} \\
25078 &  51 &   87 &  498 & 0.61 & -120 & 1 & 0.270 &  30.0 &  11.8 &  4.9 &   27 & F & \object{A 1650}    \\
39914 &  63 &   69 &  446 & 0.65 &  148 & 1 & 0.051 &  27.1 &   8.1 &  3.8 &   99 & S & \object{A 2089}   \\
50129 &  52 &   61 &  449 & 0.51 & -129 & 1 & 0.326 &  22.3 &   7.4 &  2.9 &  220 & F & \object{A 1775}   \\
58604 &  58 &   39 &  528 & 0.42 & -201 & 1 & 0.504 &  16.1 &   3.9 &  1.2 &    0 & --- & ---  \\
67116 &  80 &  114 &  651 & 0.44 & -183 & 1 & 0.094 &  36.2 &  10.5 &  3.8 &   27 & F & \object{A 1809}   \\
\label{tab:unicl}  
\end{tabular}\\
\tablefoot{                                                                                 
Columns are as in Table~\ref{tab:lumcl}.
}
\end{table*}

\begin{table*}[ht]
\caption{Data on multimodal clusters}
\begin{tabular}{rrrrrrrrrrrrrr} 
\hline\hline  
(1)&(2)&(3)&(4)&(5)& (6)&(7)&(8)&(9)& (10) & (11)&(12)& (13)& (14) \\      
\hline 
 ID   &$N_{\mathrm{gal}}$ & $L_{\mathrm{tot}}$ & $\sigma$ & $r_{\mathrm{vir}}$ &
 $V_{\mathrm{pec}}$ & $N_{\mathrm{comp}}$ & $p_{\mathrm{\Delta}}$& $D4$ & $D8$ & $D16$ & $ID_{\mathrm{scl}}$ & Type& Abell ID\\
 & & $10^{10} h^{-2} L_{\sun}$&$km~s^{-1}$&$h^{-1}$ Mpc& $km~s^{-1}$ &&&&&&& \\
\hline
  880 &  57 &  101 &  411 & 0.84 &    400 & 3 & 0.002           &  26.6 &  6.9 & 2.8 &    27 & F & ---\\
 4122 &  88 &   68 &  963 & 0.49 &  -1091 & 3 & $< 10^{-4}$ &  19.8 &  5.6 & 2.7 &   515 & S & \object{A 1291}\\
28387 &  88 &  121 &  481 & 0.66 &   -167 & 4 & $< 10^{-4}$ &  28.5 &  8.4 & 3.4 &   211 & F & --- \\
34726 & 145 &  121 &  506 & 0.74 &    -56 & 4 & 0.001 &            32.1 &  8.4 & 2.8 &   352 & S & \object{A 2028},\object{A 2033},\object{A 2040}  \\
58305 & 167 &  120 &  401 & 0.61 &    115 & 3 & $< 10^{-4}$ &  30.2 &  6.9 & 2.4 &   541 & S & \object{A 1983} \\
67297 &  95 &   86 &  770 & 0.45 &    123 & 2 & 0.001 &            26.5 &  5.4 & 1.5 &  1104 & S & \object{A 671} \\
68625 &  92 &  301 &  874 & 0.79 &    825 & 3 & $< 10^{-4}$ &  72.6 & 20.0 & 6.3 &     3 & S & \object{A 2069} \\
73088 & 141 &  184 &  631 & 0.71 &   -224 & 3 & $< 10^{-4}$ &  47.7 & 11.5 & 3.2 &    92 & S & \object{A 1904} \\
\label{tab:multicl}  
\end{tabular}\\
\tablefoot{                                                                                 
Columns are as in Table~\ref{tab:lumcl}.
}
\end{table*}


\end{appendix}

\onltab{5}{
\begin{table*}[ht]
\caption{Data on clusters}
\begin{tabular}{rrrrrrrrrrrrr} 
\hline\hline  
(1)&(2)&(3)&(4)&(5)& (6)&(7)&(8)&(9)& (10) & (11)&(12) & (13)\\      
\hline 
 ID   &$N_{\mathrm{gal}}$ & $L_{\mathrm{tot}}$ & $\sigma$ & $r_{\mathrm{vir}}$ &
 $V_{\mathrm{pec}}$ & $N_{\mathrm{comp}}$ & $p_{\mathrm{\Delta}}$& $D4$ & $D8$ & $D16$ & $ID_{\mathrm{scl}}$ & Abell ID\\
 & & $10^{10} h^{-2} L_{\sun}$&$km~s^{-1}$&$h^{-1}$ Mpc& $km~s^{-1}$ &&&&& \\
\hline
   18 &  87 &  110 &  513 & 0.73 &    387 & 2 & $< 10^{-4}$ &     25.7 &     6.8 &     2.7 &    211  & ---   \\             
  323 &  67 &   73 &  276 & 0.67 &    237 & 3 & 0.180           &     18.2 &     4.3 &     1.5 &      0  & ---   \\             
  608 &  60 &  132 &  532 & 0.66 &   -228 & 1 & 0.884           &     33.4 &     7.6 &     2.7 &    569  & A2175 \\              
  748 &  79 &   93 &  748 & 0.43 &    613 & 1 & 0.007           &     37.1 &     8.2 &     2.8 &    319  & A1066 \\              
  793 & 122 &   63 &  515 & 0.56 &   -384 & 3 & $< 10^{-4}$ &     19.4 &     4.4 &     2.1 &      0  & A2107 \\              
  880 &  57 &  101 &  411 & 0.84 &    400 & 3 & 0.002           &     26.6 &     6.9 &     2.8 &     27  & ---  \\              
  914 & 119 &  227 &  657 & 0.83 &   -704 & 5 & $< 10^{-4}$ &     49.5 &    13.2 &     4.0 &     27  & A1750 \\              
 1469 &  56 &   99 &  418 & 0.69 &   -342 & 2 & 0.198           &     32.3 &     9.1 &     3.2 &    126  &  A933 \\              
 1573 &  57 &   35 &  744 & 0.25 &   1228 & 3 & 0.334 &             -999   &  -999   &  -999   &      0  &  ---  \\             
 1944 &  60 &   47 &  440 & 0.46 &   -122 & 3 & 0.020 &               16.4 &     4.2 &     1.7 &      0  & ---   \\             
 2067 &  62 &   66 &  574 & 0.44 &   -788 & 2 & 0.007 &               21.6 &     5.0 &     1.5 &      0  & ---   \\             
 3714 &  82 &   54 &  344 & 0.60 &     -6 & 2 & $< 10^{-4}$ &     19.4 &     5.2 &     1.5 &      0  &  A1139 \\             
 4122 &  88 &   68 &  963 & 0.49 &  -1091 & 3 & $< 10^{-4}$ &     19.8 &     5.6 &     2.7 &    515  &  A1291 \\             
 4713 &  80 &   80 &  637 & 0.43 &    -78 & 3 & 0.003 &               25.6 &     7.6 &     2.9 &    515  & A1377 \\              
 4744 &  71 &   88 &  492 & 0.62 &    147 & 4 & 0.001 &               25.5 &     6.7 &     2.6 &    499  & A1238 \\              
 4992 &  68 &  157 &  619 & 0.65 &    303 & 2 & 0.048 &               48.9 &    10.1 &     3.3 &    103  &  A2048  \\            
 5217 &  89 &   94 &  577 & 0.61 &    -19 & 1 & 0.216 &               27.3 &     8.0 &     3.2 &    211  & A1436 \\              
 7102 &  54 &   33 &  465 & 0.46 &    326 & 2 & 0.022 &               14.1 &     4.8 &     2.2 &   1238  & ---   \\             
 7932 &  50 &   99 &  413 & 0.55 &   -111 & 2 & 0.089 &               29.8 &     8.5 &     2.9 &    124  & A724  \\              
 9029 &  78 &   99 &  312 & 0.72 &   -341 & 2 & 0.094 &               28.4 &     7.4 &     2.1 &    491  & A865 \\               
 9350 &  89 &   55 &  480 & 0.65 &   -129 & 4 & 0.020 &               18.5 &     7.3 &     2.7 &    352  & A2055, A2063 \\        
 9985 &  58 &   41 &  384 & 0.43 &     57 & 2 & 0.167 &               18.6 &     6.4 &     2.4 &    782  & ---  \\              
10438 &  65 &   45 &  395 & 0.46 &   -161 & 2 & 0.008 &               15.4 &     4.9 &     2.1 &    782  & ---   \\             
11015 &  52 &   45 &  303 & 0.46 &    140 & 8 & 0.037 &               17.9 &     4.5 &     1.5 &      0  & ---   \\             
11474 &  51 &   36 &  306 & 0.49 &   -267 & 5 & 0.001 &               11.5 &     3.1 &     1.1 &      0  & ---   \\             
11683 &  54 &   52 &  342 & 0.46 &    -21 & 1 & 0.064 &               17.3 &     3.5 &     1.8 &      0  & A1507 \\              
12508 &  93 &   90 &  385 & 0.77 &    -78 & 2 & $< 10^{-4}$ &     21.2 &     6.0 &     1.7 &    850  & A2169 \\              
12540 & 103 &  107 &  764 & 0.52 &    477 & 2 & 0.345 &               35.3 &     9.8 &     3.4 &    143  & A1767 \\              
13216 &  57 &   48 &  400 & 0.54 &    261 & 3 & $< 10^{-4}$ &     16.3 &     3.6 &     1.3 &      0  & ---   \\             
13347 &  50 &   64 &  479 & 0.48 &    369 & 2 & 0.007 &               19.8 &     4.9 &     1.5 &      0  & A1003 \\              
13408 &  58 &   62 &  282 & 0.73 &    114 & 2 & 0.752 &               18.8 &     6.0 &     1.7 &    849  & A2149  \\             
16094 &  71 &  113 &  735 & 0.50 &    -47 & 3 & 0.021 &               37.1 &     8.8 &     2.4 &    218  & A1691 \\              
16309 &  69 &  151 &  878 & 0.46 &     84 & 4 & 0.004 &               43.4 &     9.1 &     2.6 &     59  & A2244 \\              
16350 &  65 &  135 &  859 & 0.36 &   -626 & 2 & $< 10^{-4}$ &     43.4 &     8.6 &     2.9 &    233  & A2245   \\            
17210 &  72 &  120 &  872 & 0.47 &  -1308 & 2 & 0.007 &               36.0 &     8.4 &     2.7 &    233  & A2249 \\              
18029 &  53 &   37 &  597 & 0.41 &   -625 & 3 & $< 10^{-4}$ &     12.4 &     2.8 &     1.1 &      0  & A1781 \\              
18048 &  78 &  121 &  596 & 0.61 &     28 & 2 & 0.010 &               39.7 &    12.0 &     4.2 &     54  & ---   \\             
20159 &  52 &   59 &  517 & 0.36 &   -300 & 2 & 0.011 &               20.6 &     4.1 &     1.3 &      0  & A1026, A1035 \\        
20419 &  58 &   44 &  424 & 0.59 &    201 & 3 & 0.004 &               15.1 &     3.6 &     1.7 &      0  & A1749 \\              
20514 &  56 &   33 &  317 & 0.42 &   -390 & 4 & 0.436 &               12.5 &     3.9 &     1.7 &      0  & ---   \\             
21573 &  50 &   60 &  364 & 0.49 &    -32 & 2 & 0.488 &               22.4 &     6.6 &     2.3 &    797  & ---   \\             
22572 &  77 &   75 &  533 & 0.43 &    -33 & 2 & $< 10^{-4}$ &     21.1 &     4.5 &     1.5 &      0  & A1169 \\              
23374 & 114 &  100 &  662 & 0.72 &   1226 & 2 & 0.001 &               27.6 &     8.3 &     3.4 &    219  & A1795, A1818 \\        
23524 &  50 &   63 &  304 & 0.67 &    -68 & 2 & 0.166 &               19.8 &     7.5 &     2.9 &    220  & ---   \\             
24554 &  50 &   83 &  658 & 0.58 &    107 & 3 & $< 10^{-4}$ &     27.9 &    10.1 &     4.4 &     27  & A1620 \\              
24604 &  50 &   59 &  857 & 0.50 &   -256 & 2 & 0.001 &               21.2 &     6.8 &     3.0 &    220  & A1831\\               
24829 &  77 &  126 &  534 & 0.66 &   -107 & 3 & 0.003 &               36.8 &    11.5 &     4.3 &     27  & A1663\\               
25078 &  51 &   87 &  498 & 0.61 &   -120 & 1 & 0.270 &               30.0 &    11.8 &     4.9 &     27  & A1650 \\              
28272 &  51 &   40 &  355 & 0.50 &    464 & 1 & 0.032 &               13.0 &     4.8 &     1.3 &    868  & ---   \\             
28387 &  88 &  121 &  481 & 0.66 &   -167 & 4 & $< 10^{-4}$ &     28.5 &     8.4 &     3.4 &    211  & ---   \\             
28508 &  58 &   74 &  494 & 0.40 &    478 & 1 & 0.482 &               24.8 &     8.0 &     3.3 &    211  & A1270 \\              
28986 &  66 &   73 &  398 & 0.69 &   -205 & 5 & 0.002 &               19.7 &     6.4 &     2.8 &    360  & A2092 \\              
29348 &  75 &  119 &  418 & 0.69 &      9 & 4 & 0.001 &               33.7 &     8.7 &     3.1 &     99  &  ---  \\             
29350 &  55 &   92 &  334 & 0.64 &   -309 & 2 & 0.325 &               29.9 &     8.0 &     3.1 &     99  & ---   \\             
29587 & 207 &  365 &  740 & 0.87 &    334 & 3 & $< 10^{-4}$ &     79.3 &    21.3 &     5.8 &      1  & A2142 \\              
29744 &  53 &   96 &  396 & 0.76 &   -557 & 3 & 0.004 &               24.0 &     5.9 &     2.1 &    870  & A1939 \\              
30391 &  68 &   93 &  271 & 0.73 &    279 & 3 & $< 10^{-4}$ &     20.4 &     4.6 &     1.4 &      0  & ---   \\             
32006 &  71 &   72 &  420 & 0.53 &     11 & 2 & $< 10^{-4}$ &     19.9 &     5.6 &     2.7 &    211  & A1396, A1400 \\        
32663 &  51 &   61 &  435 & 0.73 &     96 & 1 & 0.008 &               19.3 &     4.8 &     2.0 &      0  & ---   \\             
32909 &  79 &   69 &  560 & 0.56 &   -223 & 3 & 0.006 &               22.8 &     5.3 &     1.8 &   1244  & A2022 \\              
32976 &  64 &   72 &  510 & 0.68 &    157 & 2 & 0.053 &               21.7 &     4.8 &     1.4 &      0  & A1808 \\              
33082 &  77 &   79 &  382 & 0.52 &   -137 & 2 & 0.120 &               25.8 &     6.9 &     2.9 &    211  & A1383, A1396, A1400 \\  
33739 &  79 &   65 &  517 & 0.49 &   -217 & 4 & 0.003 &               21.5 &     5.0 &     2.0 &      0  & A1890 \\              
33851 & 138 &   74 &  354 & 0.77 &    472 & 3 & 0.057 &               18.4 &     4.8 &     1.9 &      0  & ---   \\             
34513 &  53 &   94 &  426 & 0.75 &    176 & 2 & 0.004 &               22.4 &     5.0 &     2.1 &   1295  & A2020 \\              
\label{tab:cldata1}  
\end{tabular}\\
\tablefoot{                                                                                 
Columns are as follows:
1: ID of a cluster;
2: the number of galaxies in the cluster, $N_{\mathrm{gal}}$;
3: total luminosity of  the cluster;
4: rms velocity of the cluster;
5: virial radius of the cluster;
6:  peculiar velocity of the main galaxy;
7: the number of components in the cluster, $N_{\mathrm{comp}}$;
8: p-value of the $\Delta$ test;
9--11: environmental density around the cluster, at smoothing lengths 4, 8, and 16 \Mpc
(in units of the mean density; -999 denotes cluster at the edge of a survey where
the density cannot be calculated);
12: ID of the supercluster where the cluster resides;
13: Abell ID of the cluster.
}
\end{table*}
}

\onltab{5}{
\begin{table*}[ht]
\caption{...continued}
\begin{tabular}{rrrrrrrrrrrrr} 
\hline\hline  
(1)&(2)&(3)&(4)&(5)& (6)&(7)&(8)&(9)& (10) & (11)&(12)& (13) \\      
\hline 
 ID   &$N_{\mathrm{gal}}$ & $L_{\mathrm{tot}}$ & $\sigma$ & $r_{\mathrm{vir}}$ &
 $V_{\mathrm{pec}}$ & $N_{\mathrm{comp}}$ & $p_{\mathrm{\Delta}}$& $D4$ & $D8$ & $D16$ & $ID_{\mathrm{scl}}$ & Abell ID\\
 & & $10^{10} h^{-2} L_{\sun}$&$km~s^{-1}$&$h^{-1}$ Mpc& $km~s^{-1}$ &&&&& \\
\hline
34726 & 145 &  121 &  506 & 0.74 &    -56 & 4 & 0.001 &               32.1 &     8.4 &     2.8 &    352   & A2028, A2033, A2040 \\      
34727 & 256 &  351 &  825 & 1.25 &   -290 & 5 & $< 10^{-4}$ &     51.3 &    16.1 &     5.1 &      7   & A2028, A2029, A2033, A2040\\ 
35037 &  79 &   76 &  691 & 0.58 &    748 & 4 & 0.028 &               22.3 &     5.8 &     2.4 &    865   & A2122, A2124 \\            
36861 &  66 &  110 &  493 & 0.84 &    379 & 3 & $< 10^{-4}$ &     23.7 &     5.6 &     1.6 &   1192   & A1616 \\                  
38087 & 169 &  209 &  541 & 0.84 &    844 & 2 & 0.031 &               51.1 &    13.9 &     3.7 &     24   & A1173, A1187, A1190, A1203 \\
39489 & 166 &  188 & 1061 & 0.72 &   -521 & 3 & $< 10^{-4}$ &     43.5 &    10.7 &     4.0 &     99   & A2056, A2065\\             
39752 & 108 &  116 &  514 & 0.75 &    146 & 2 & 0.004 &               30.2 &     7.5 &     2.7 &    360   & A2073, A2079 \\            
39914 &  63 &   69 &  446 & 0.65 &    148 & 1 & 0.051 &               27.1 &     8.1 &     3.8 &     99   & A2089  \\                 
40520 &  52 &   56 &  486 & 0.67 &   -348 & 2 & $< 10^{-4}$ &     22.8 &     9.0 &     3.1 &     89   & A1569 \\                  
40870 & 118 &  153 &  717 & 0.69 &   -304 & 2 & 0.001 &               40.6 &     9.5 &     3.0 &     89   & ---  \\                  
42481 &  57 &   55 &  354 & 0.53 &   -184 & 3 & $< 10^{-4}$ &     18.5 &     5.7 &     1.6 &    868   & ---  \\                  
43336 &  68 &   73 &  471 & 0.60 &    -97 & 1 & 0.003 &               22.4 &     6.7 &     2.3 &    532   & ---  \\                  
43545 &  51 &   42 &  577 & 0.51 &     81 & 1 & 0.004 &               13.6 &     4.2 &     1.5 &      0   & A602 \\                   
43966 &  74 &   76 &  613 & 0.52 &    -10 & 2 & 0.155 &               29.2 &     7.7 &     2.3 &    344   & A1668, A1669 \\            
44471 & 113 &   90 &  464 & 0.55 &   -411 & 4 & $< 10^{-4}$ &     25.3 &     6.1 &     1.9 &    793   & A1185, A1213 \\            
47492 &  74 &   57 &  458 & 0.56 &    184 & 3 & 0.307 &               18.6 &     6.4 &     2.3 &    782   & ---  \\                  
48448 &  55 &   41 &  340 & 0.54 &    192 & 2 & 0.085 &               12.8 &     4.6 &     2.1 &      0   & A784 \\                   
50129 &  52 &   61 &  449 & 0.51 &   -129 & 1 & 0.326 &               22.3 &     7.4 &     2.9 &    220   & A1775 \\                  
50631 & 101 &   86 &  636 & 0.55 &   -144 & 2 & 0.032 &               22.7 &     4.5 &     1.5 &      0   & ---  \\                  
50647 &  52 &   62 &  525 & 0.52 &   -195 & 2 & 0.259 &               19.6 &     4.7 &     2.1 &      0   & ---  \\                  
50657 &  55 &   49 &  555 & 0.42 &    427 & 1 & 0.078 &               13.5 &     2.7 &     1.0 &      0   & ---  \\                  
52913 &  67 &  116 &  368 & 0.77 &   -287 & 3 & 0.011 &               24.7 &     5.9 &     2.2 &    721   & A690 \\                   
56571 &  55 &  139 &  457 & 0.60 &   -368 & 4 & 0.016 &               36.2 &     6.9 &     2.0 &    543   & A1999, A2000 \\            
57317 & 118 &  105 &  516 & 0.59 &   -712 & 2 & $< 10^{-4}$ &     37.9 &     8.3 &     2.4 &    349   & A1913 \\                  
58101 & 105 &  122 &  614 & 0.92 &     26 & 4 & $< 10^{-4}$ &     27.3 &     7.4 &     2.6 &    499   & A1205\\                   
58305 & 167 &  120 &  401 & 0.61 &    115 & 3 & $< 10^{-4}$ &     30.2 &     6.9 &     2.4 &    541   & A1983\\                   
58323 &  64 &   37 &  393 & 0.40 &    -56 & 3 & 0.259 &               12.3 &     3.1 &     1.4 &      0   & --- \\                   
58604 &  58 &   39 &  528 & 0.42 &   -201 & 1 & 0.504 &               16.1 &     3.9 &     1.2 &      0   & ---  \\                  
59794 &  90 &  141 &  650 & 0.75 &   -881 & 2 & $< 10^{-4}$ &     36.7 &     8.9 &     3.2 &    214   & A1552 \\                  
60539 & 107 &  136 &  830 & 0.55 &    -24 & 1 & 0.021 &               47.8 &    14.1 &     5.0 &     19   & A1516  \\                 
61613 &  77 &  138 &  517 & 1.06 &    100 & 4 & $< 10^{-4}$ &     25.7 &     7.4 &     2.9 &     54   & A1358 \\                  
62138 & 124 &  127 &  456 & 0.78 &    118 & 6 & $< 10^{-4}$ &     26.8 &     5.7 &     1.7 &   1247   & A1991 \\                  
63361 &  72 &  103 &  646 & 0.64 &   -175 & 1 & 0.004 &               34.4 &    10.0 &     3.6 &     27   & A1773 \\                  
63757 &  87 &  155 &  653 & 0.65 &   -159 & 4 & $< 10^{-4}$ &     41.5 &     9.4 &     3.2 &    214   & A1541 \\                  
63949 &  80 &  112 &  661 & 0.55 &   -843 & 3 & 0.001 &               33.8 &    10.1 &     3.9 &     19   & A1424 \\                  
64635 & 109 &   96 &  489 & 0.73 &   -347 & 3 & 0.001 &               25.9 &     7.5 &     2.4 &    540   & ---  \\                  
64702 &  64 &   84 &  539 & 0.74 &    575 & 2 & 0.001 &               25.8 &    11.7 &     4.6 &     19   & A1516 \\                  
67116 &  80 &  114 &  651 & 0.44 &   -183 & 1 & 0.094 &               36.2 &    10.5 &     3.8 &     27   & A1809 \\                  
67297 &  95 &   86 &  770 & 0.45 &    123 & 2 & 0.001 &               26.5 &     5.4 &     1.5 &   1104   & A671 \\                   
68376 & 106 &  157 &  671 & 0.51 &    327 & 3 & 0.205 &               46.4 &    10.8 &     4.1 &     99   & A2061, A2067 \\            
68625 &  92 &  301 &  874 & 0.79 &    825 & 3 & $< 10^{-4}$ &     72.6 &    20.0 &     6.3 &      3   & A2069 \\                  
73088 & 141 &  184 &  631 & 0.71 &   -224 & 3 & $< 10^{-4}$ &     47.7 &    11.5 &     3.2 &     92   & A1904 \\                  
73420 &  68 &  105 &  555 & 0.73 &    221 & 3 & $< 10^{-4}$ &     31.8 &     8.2 &     2.6 &    298   & A628 \\                   
74783 &  65 &   46 &  401 & 0.48 &    -55 & 3 & 0.138 &               14.5 &     3.7 &     1.4 &      0   & ---  \\                  
\label{tab:cldata2}                                                                                                
\end{tabular}\\                                                                                                    
\tablefoot{                                                                                                        
Columns are as follows:                                                                                            
Columns are as follows:
1: ID of a cluster;
2: the number of galaxies in the cluster, $N_{\mathrm{gal}}$;
3: total luminosity of  the cluster;
4: rms velocity of the cluster;
5: virial radius of the cluster;
6:  peculiar velocity of the main galaxy;
7: the number of components in the cluster, $N_{\mathrm{comp}}$;
8: p-value of the $\Delta$ test;
9--11: environmental density around the cluster, at smoothing lengths 4, 8, and 16 \Mpc
(in units of the mean density; -999 denotes cluster at the edge of a survey where
the density cannot be calculated);
12: ID of the supercluster where the cluster resides;
13: Abell ID of the cluster.
}                                                                                                                  
\end{table*}                                                                                                       
}

\onltab{6}{
\begin{table*}[ht]
\caption{Data on superclusters}
\begin{tabular}{rrrrrrrrrrrrr} 
\hline\hline  
(1)&(2)&(3)&(4)&(5)& (6)&(7)&(8)&(9)&(10)& (11) & (12)& (13)\\      
\hline 
 ID(long)   &ID(DR8) & ID(DR7)&  $N_{\mathrm{gal}}$ & $Dist.$&$L_{\mathrm{tot}}$ & $D_{\mathrm{peak}}$ &
$V_3$ & $K_1$ & $K_2$ & $K_1/K_2$ & Type & ID(E01)\\
 & & & & $h^{-1}$ Mpc&$10^{10} h^{-2} L_{\sun}$& &&&&&&\\

\hline
239+027+0091 &   1 &    1 & 1041& 264& 1809   &  22.2   &   2.5 &  0.07  &  0.17 &   0.41  & F & 162    \\
231+030+0117 &   3 &    5 & 1191& 336& 3694   &  20.6   &   8.5 &  0.12  &  0.17 &   0.74  & S & 157     \\
227+006+0078 &   7 &   11 & 1217& 233& 1675   &  17.0   &   3.0 &  0.04  &  0.08 &   0.56  & S & 154    \\
184+003+0077 &  19 &   24 & 2060& 231& 2919   &  15.0   &   9.0 &  0.11  &  0.34 &   0.33  & S & 111    \\
167+040+0078 &  24 &   38 &  580& 225&  751   &  14.6   &   2.0 &  0.02  &  0.03 &   0.86  & S & 95    \\
202-001+0084 &  27 &   61 & 3222& 256& 5163   &  14.0   &  14.5 &  0.12  &  0.46 &   0.26  & F & 126    \\
173+014+0082 &  54 &   55 & 1341& 241& 2064   &  12.4   &   5.5 &  0.09  &  0.16 &   0.56  & S & 111    \\
250+027+0102 &  59 &   64 &  656& 301& 1563   &  12.8   &   5.0 &  0.08  &  0.22 &   0.36  & F & 164    \\
189+017+0071 &  89 &  136 &  515& 212&  610   &  11.5   &   2.0 &  0.03  &  0.01 &   1.98  & S & 271    \\
215+048+0071 &  92 &   87 &  527& 213&  690   &  11.8   &   2.5 &  0.05  &  0.12 &   0.40  & S & ---    \\
230+027+0070 &  99 &   94 & 2047& 215& 2874   &  11.5   &  10.0 &  0.11  &  0.43 &   0.26  & S & 158    \\
229+006+0102 & 103 &  152 &  425& 302& 1004   &  11.0   &   3.0 &  0.05  &  0.10 &   0.50  & F & 160    \\
135+038+0094 & 124 &  195 &  273& 281&  548   &  10.5   &   2.0 &  0.02  &  0.03 &   0.62  & S & 249    \\
152-000+0096 & 126 &  198 &  495& 285& 1001   &  10.1   &   4.0 &  0.05  &  0.13 &   0.37  & F & 82    \\
203+059+0072 & 143 &  228 &  668& 211&  753   &  10.1   &   2.0 &  0.04  &  0.03 &   1.27  & F & 133    \\
172+054+0071 & 211 &  336 & 1439& 207& 1618   &   9.2   &   7.0 &  0.10  &  0.28 &   0.36  & F & 109    \\
187+008+0090 & 214 &  223 &  735& 268& 1218   &   9.7   &   6.0 &  0.08  &  0.30 &   0.28  & S & 111    \\
197+039+0073 & 218 &  344 &  272& 215&  337   &   9.1   &   1.0 &  0.00  & -0.00 & -20.02  & S & 274    \\
207+026+0067 & 219 &  349 &  985& 187& 1007   &   9.2   &   4.0 &  0.07  &  0.11 &   0.61  & S & 138    \\
207+028+0077 & 220 &  351 &  603& 226&  768   &   9.0   &   4.0 &  0.05  &  0.10 &   0.47  & F & 138    \\
255+033+0086 & 233 &  376 &  474& 259&  790   &   9.0   &   4.0 &  0.04  &  0.04 &   0.96  & F & 167    \\
122+035+0084 & 298 &  311 &  246& 246&  345   &   8.6   &   1.0 &  0.02  &  0.00 &   4.27  & S & 75    \\
159+004+0069 & 319 &  503 &  245& 207&  296   &   8.3   &   2.0 &  0.02  &  0.02 &   0.97  & S & 91    \\
195+019+0064 & 344 &  538 &  264& 192&  290   &   8.1   &   1.0 & -0.00  &  0.01 &  -0.14  & S & 273    \\
216+016+0051 & 349 &  548 &  335& 159&  284   &   8.5   &   1.0 &  0.01  &  0.01 &   0.97  & S & 143    \\
227+007+0045 & 352 &  550 &  519& 135&  379   &   8.7   &   1.0 &  0.02  &  0.01 &   1.46  & S & 154    \\
232+029+0066 & 360 &  362 &  311& 196&  330   &   8.4   &   2.0 &  0.00  & -0.01 &  -0.54  & S & 158    \\
146+043+0072 & 491 &  500 &  199& 217&  256   &   7.8   &   1.0 &  0.00  & -0.01 &  -0.15  & S & ---    \\
168+002+0077 & 499 &  512 &  408& 228&  517   &   7.8   &   3.0 &  0.05  &  0.08 &   0.67  & F & 91    \\
176+055+0052 & 515 &  525 &  457& 155&  390   &   7.9   &   3.0 &  0.02  & -0.00 & -11.10  & S & 109    \\
208+046+0062 & 532 &  543 &  297& 189&  293   &   7.4   &   2.0 &  0.01  &  0.02 &   0.49  & F & ---    \\
214+002+0053 & 540 &  549 &  422& 163&  358   &   7.9   &   1.5 &  0.02  &  0.05 &   0.49  & S & ---    \\
223+016+0045 & 541 &  849 &  299& 135&  214   &   7.2   &   1.0 &  0.01  & -0.01 &  -0.88  & S & ---    \\
223+054+0098 & 543 &  552 &   80& 294&  186   &   7.0   &   1.0 &  0.01  & -0.04 &  -0.40  & S & 147    \\
245+029+0096 & 569 &  578 &  198& 285&  417   &   7.9   &   2.0 &  0.04  &  0.05 &   0.79  & S & 164    \\
129+028+0079 & 721 & 1094 &   94& 237&  145   &   6.2   &   1.0 & -0.00  & -0.01 &   0.22  & S & 76    \\
151+054+0047 & 782 &  779 &  652& 139&  465   &   6.8   &   3.0 &  0.05  &  0.14 &   0.38  & S & ---    \\
169+029+0046 & 793 & 1184 &  211& 142&  156   &   6.3   &   1.0 & -0.01  &  0.01 &  -1.24  & S & 93    \\
176+015+0069 & 797 &  794 &  160& 205&  203   &   6.7   &   1.0 &  0.00  &  0.01 &   0.22  & S & ---    \\
240+053+0065 & 849 &  857 &  215& 194&  242   &   6.5   &   1.0 &  0.00  & -0.02 &  -0.03  & F & 162    \\
244+049+0057 & 850 & 1258 &  135& 171&  128   &   6.2   &   1.0 & -0.01  & -0.02 &   0.63  & S & 162    \\
234+036+0065 & 865 & 1278 &  161& 197&  170   &   6.0   &   2.0 &  0.02  & -0.01 &  -1.68  & S & 158    \\
246+014+0050 & 868 & 1283 &  239& 153&  201   &   6.4   &   1.5 & -0.00  & -0.00 &   0.98  & S & ---    \\
219+024+0087 & 870 & 1284 &   76& 261&  143   &   6.1   &   1.0 &  0.00  & -0.01 &  -0.14  & S & ---    \\
127+030+0051 &1104 & 1683 &  127& 150&  110   &   5.6   &   1.0 & -0.01  & -0.02 &   0.68  & S & ---    \\
191+054+0085 &1192 & 1208 &   79& 248&  134   &   5.9   &   1.0 & -0.01  & -0.00 &   1.41  & S & ---    \\
220+010+0051 &1238 & 1255 &  133& 156&   97   &   5.5   &   1.0 &  0.00  & -0.00 &  -0.28  & F & ---    \\
226+028+0057 &1244 & 1262 &  101& 174&   97   &   5.4   &   1.0 & -0.01  & -0.02 &   0.76  & S & 152    \\
223+018+0059 &1247 & 1265 &  147& 176&  149   &   5.9   &   1.0 & -0.00  & -0.03 &   0.28  & F & ---    \\
226+007+0092 &1295 & 1311 &  110& 272&  200   &   5.4   &   2.0 &  0.02  &  0.04 &   0.44  & F & ---    \\
\label{tab:scl}  
\end{tabular}\\
\tablefoot{                                                                                 
The columns are as follows:
1: Long ID of a supercluster AAA+BBB+ZZZZ, 
where AAA is R.A., +/-BBB is Dec. (in degrees), and ZZZZ is 1000$z$;
2: ID(DR8) of a supercluster from the SDSS DR8 superclusters catalogue;
3: ID of a supercluster from the SDSS DR7 superclusters catalogue 
\citep{2010arXiv1012.1989J};
4: the number of galaxies in the supercluster, $N_{\mbox{gal}}$;
5: the distance of a supercluster;
6: the total weighted luminosity of galaxies in the supercluster, $L_{\mbox{tot}}$;
7: the density at the density maximum, $d_{\mbox{peak}}$, in units of mean density;
8: the maximum value of the fourth Minkowski functional,
($V_{3,\mathrm{max}}$ (clumpiness), for the supercluster;
9 -- 11: the shapefinders $K_1$ (planarity) and $K_2$ (filamentarity), and 
the ratio of the shapefinders $K_1/K_2$ for the full supercluster;
12: Morphological type of a supercluster
13: ID(E01): the supercluster ID  in  the catalogue by \citet{e2001}. SCl~160 -- the Hercules
supercluster, SCl~111 and SCl~126 -- members of the Sloan Great Wall, SCl~158 -- the Corona 
Borealis supercluster, SCl~138 -- the Bootes supercluster, SCl~336 -- the Ursa 
 Major supercluster.
}
\end{table*}
}
                                                                                                                   
                                                                                                                   
\end{document}